\documentclass[twocolumn,aps,tightenlines,superscriptaddress,showpacs,nofootinbib,floatfix]{revtex4-2}
\usepackage{amsmath,amsfonts,amssymb,latexsym}
\usepackage{hyperref}
\usepackage{epsfig,bm,feynmf}
\usepackage{graphicx}
\usepackage{amsmath}
\usepackage{dcolumn}
\usepackage[sort&compress]{natbib}
\usepackage{grffile,epstopdf}
\usepackage[usenames,dvipsnames,svgnames,table]{xcolor}
\usepackage{hhline}
\usepackage{physics}
\usepackage[T1]{fontenc}

\hypersetup{
  colorlinks,
  unicode=True,
  linkcolor=Blue,
  citecolor=Blue,
  urlcolor=Blue,
  pdftitle={Extraction of the microscopic properties of quasi-particles using DNN},
  pdfauthor={Olga Soloveva},
  pdfsubject={HEP}
}
\graphicspath{ {pics/} {NN/} }
\newcommand{\nB}{\langle n_{B} \rangle }
\newcommand{\nS}{\langle n_{S} \rangle }
\newcommand{\nuu}{\langle n_{u} \rangle }
\newcommand{\nd}{\langle n_{d} \rangle }

\newcommand{\ns}{\langle n_{s} \rangle }
\newcommand{\di}{\mathrm{d}}
\begin{document}

\title{Extraction of the microscopic properties of quasi-particles using deep neural networks}

%%%%%%%%%%%%%%%%%%%%% Authors 
\author{Olga Soloveva}
\email{soloveva@itp.uni-frankfurt.de}
\affiliation{Helmholtz Research Academy Hesse for FAIR (HFHF), GSI Helmholtz Center for Heavy Ion Physics, Campus Frankfurt, 60438 Frankfurt, Germany}
\affiliation{Institut f\"ur Theoretische Physik, Johann Wolfgang Goethe-Universit\"at, Max-von-Laue-Str.\ 1, D-60438 Frankfurt am Main, Germany}
\author{Andrea Palermo}
\email{andrea.palermo@stonybrook.edu}
\affiliation{Center for Nuclear Theory, Department of Physics and Astronomy,
Stony Brook University, Stony Brook, New York 11794–3800, USA}
\author{Elena Bratkovskaya}
\affiliation{Helmholtz Research Academy Hesse for FAIR (HFHF), GSI Helmholtz Center for Heavy Ion Physics, Campus Frankfurt, 60438 Frankfurt, Germany}
\affiliation{Institut f\"ur Theoretische Physik, Johann Wolfgang Goethe-Universit\"at, Max-von-Laue-Str.\ 1, D-60438 Frankfurt am Main, Germany}
\affiliation{GSI Helmholtzzentrum f\"ur Schwerionenforschung GmbH,Planckstrasse 1, D-64291 Darmstadt, Germany}
%%%%%%%%%%%%%%%%%%%%% 

%%%%%%%%%%%%%%%%%%%% Abstract %%%%%%%%%%%%%%%%%%%%%
\date{\today}
\begin{abstract}
We use deep neural networks (DNN) to obtain the microscopic characteristics of partons in terms of dynamical degrees of freedom on the basis of an off-shell quasiparticle description. We aim to infer masses and widths of quasi-gluons, up/down, and strange quarks using constraints on the macroscopic thermodynamic observables obtained by the first-principles calculations lattice QCD. In this work, we use 3 independent dimensionless thermodynamic observables from lQCD for minimization.  First, we train our DNN using the DQPM (Dynamical QuasiParticle Model) Ansatz for the masses and widths. 
Furthermore, we use the DNN capabilities to generalize this Ansatz, to evaluate which quasiparticle characteristics are desirable to describe different thermodynamic functions simultaneously. To evaluate consistently the microscopic properties obtained by the DNN  in the case of off-shell quarks and gluons, we compute transport coefficients using the spectral function within Kubo-Zubarev formalism in different setups. In particular, we make a comprehensive comparison in the case of the dimensionless ratios of shear viscosity over entropy density $\eta/s$ and electric conductivity over temperature $\sigma_Q/T$, which provide additional constraints for the parameter generalization of the considered models.
\end{abstract}

\maketitle
%%%%%%%%%%%%%%%%%%%%%%%%%%%%%%%%%%%%%%%%%%%%%%%%%%%%%%%%%%%%%%%%%%%%%%%%
\section{Introduction}
\label{sec:Intro}
In the exploration of quark-gluon plasma (QGP) matter produced during heavy-ion collisions, a critical step is establishing the equation of state (EoS) and transport coefficients of the matter corresponding to a specific range of parameters, such as energy density, which is associated with temperature, and baryon density, associated with baryon chemical potential. Moreover, for the fully dynamical evolution of the system, it is required to consider microscopic characteristics for the partonic degrees of freedom. 
Although first-principles Lattice Quantum Chromodynamics (lQCD) simulations provide vital predictions, especially for temperatures above the critical temperature $T_c$, they are complex and challenging to execute. 
While lQCD has successfully outlined the main thermodynamic quantities, finer details of microscopic characteristics and dynamic properties, such as transport coefficients, remain somewhat elusive. These properties are crucial as they depend on the microscopic interactions among quarks and gluons. However, assessing the microscopic properties of QGP matter at finite temperatures and baryon chemical potentials from first principles is a challenging task.
To bridge this gap, it is worthwhile to investigate new methodologies that can further elucidate the established lQCD predictions, thereby enhancing our understanding of the QGP matter in these extreme conditions.
In the past decade, methods of machine learning (ML) have been developed as a powerful computational tool and novel problem-solving perspective for
physics, which offers new avenues for exploring strongly interacting QCD matter under extreme conditions such as finite $T$ and baryon chemical potential \cite{Pang:2016vdc, Huang:2018fzn, Palermo:2021jrf, OmanaKuttan:2022aml, Soma:2022vbb, Aarts:2023vsf,Ermann:2023unw,Wang:2023muv,He:2023zin}. \\

Generally, ML stands apart from conventional minimization methods by aiming for predictions instead of just fitting, thus offering enhanced adaptability for applications in physics \cite{Mehta:2018dln, Zhou:2023pti}. In the context of  Relativistic Heavy-Ion Collisions (HICs) and phenomenology of the QGP, this includes the use of novel techniques such as active learning \cite{Mroczek:2022oga}, transfer learning \cite{Liyanage:2022byj}, among other methods \cite{Albergo:2019eim,Blucher:2020mjt,Abbott:2022hkm,Caselle:2023mvh}.
In this work, we use machine learning techniques, in particular, Deep Neural Networks (DNNs) \cite{mcculloch1943logical,schmidhuber2015deep}, to gain insights into the phenomenological description of deconfined QCD matter and test the applicability of these methods to phenomenology. We confine ourselves to the case of $2+1$ flavours and colours QCD, $N_f=N_c=3$, with degenerate light quarks at vanishing quark chemical potential, $\mu_q=0$. 
To train the DNNs used in this work we use lQCD data provided by the WB Collaboration, in particular results on the entropy density $s(T)$ and baryon susceptibilities $\chi_2^B(T), \chi_2^S(T)$ \cite{Borsanyi:2013bia, Borsanyi:2012cr} $(T \leq 3T_c)$ and latest estimates for the strange susceptibility $\chi_2^S(T)$ \cite{Borsanyi:2022qlh} $(T \leq 1.7 T_c)$. 
We choose these thermodynamic quantities in order to fix consistently the strange quark, the light quark, and gluon sectors independently. The functional form of the masses and widths of quarks and gluons is inferred by multi-layer Feed-forward neural networks (FFNNs), which are trained on the entropy density, baryon, and strange susceptibilities.
The feasibility of this approach is grounded in the universal approximation theorem \cite{HORNIK1989359, pml2Book}. This theorem states that multi-layer feed-forward neural networks (FFNNs), owing to their non-linear characteristics and given a sufficient number of hidden neurons, can effectively approximate any function.

To evaluate the thermodynamic quantities required for the training of DNNs we consider an off-shell quasi-particle description of the QGP, where quasi-quarks and gluons acquire a thermal width as in the original Dynamical Quasi-Particle Model (DQPM) \cite{Peshier:2005pp, Cassing:2007nb, Cassing:2007yg, Berrehrah:2016vzw, Moreau:2019vhw, Soloveva:2020hpr, Grishmanovskii:2023gog}. 

The DQPM is based on a propagator representation with effective (anti-)quarks and gluons whose properties are defined by complex self-energies and off-shell spectral functions. 
The advantage of the DQPM compared to other QPM  is that it is by construction a 2PI (two particle irreducible) model while the other on-shell quasiparticle models are 1PI (one particle irreducible) in nature. 
This avoids the introduction of an extra ``bag constant''  often employed for on-shell quasi-particle models in order to describe the thermodynamic properties of the QGP.

Quasiparticle models are highly favored for their ease of integration into transport frameworks, which are crucial for simulating the evolution of QGP matter. Specifically, the DQPM has been incorporated into the Parton-Hadron-String Dynamics (PHSD) transport approach \cite{Cassing:2009vt, Moreau:2019vhw}. Meanwhile, the Quasiparticle Model (QPM) has been adopted in the Catania transport approach \cite{Plumari:2011mk} and \cite{Scardina:2017ipo}. Additionally, recent developments have seen the implementation of approximate QCD models, like the Nambu–Jona-Lasinio (NJL) and Polyakov–Nambu–Jona-Lasinio (PNJL) models, within the A Multi-Phase Transport (AMPT) model \cite{Sun:2020uoj}. These implementations utilize scalar and vector potentials to approximate QCD dynamics. 

In addition to analyzing thermodynamic quantities, we also explore transport coefficients, using the resulting microscopic properties, i.e. spectral functions. 
Our previous systematic comparison reveals that these transport coefficients are significantly influenced by the properties of the constituent degrees of freedom \cite{Soloveva:2021quj}. This leads to the observation that different theoretical models, despite having an almost identical EoS, can yield markedly different transport coefficients. 
The goal of this study is to explore the properties of strongly interacting quasi-particles, i.e. the $T$- dependence of masses and widths which are consistent with the lQCD thermodynamics at vanishing chemical potential, by using DNN methods. The techniques used here can be also applied at finite $\mu_b$.
Starting from the DQPM Ansatz we gradually generalize it, aiming at a simultaneous description of both the entropy density $s(T)$ and the susceptibilities $\chi_2^{B,S}(T)$. 

In general, a substantial amount of data is essential to effectively train a neural network to uncover established physics. In the phenomenology of the  QGP, these get even worse since the only available lQCD data represents a small amount of points. Therefore the simple architecture and small number of hyperparameters of FFNNs compared to CNNs and ResNets is beneficial for this task.

The structure of the paper is as follows. In Sec. \ref{sec-quasi} we briefly 
describe the framework of the off-shell Quasi-particle model including the evaluation of thermodynamic observables and transport coefficients. In Sec. \ref{sec-dnn-scheme} we describe our method to determine the microscopic quantities using DNNs, starting from the DQPM and generalizing it further in Sec. \ref{sec:dnn-gen}.
Finally, our results and conclusions are summarized in Sec.~\ref{sec:conclusions}.
%--------------------------------------------------------------------------------------
\section{ \label{sec-quasi}Framework of the off-shell quasiparticle model}

In the quasiparticle approach, the degrees of freedom are strongly interacting dynamical quasiparticles: quarks and gluons with a broad spectral function, whose `thermal' masses and widths increase with temperature. The microscopic and macroscopic properties of the off-shell quasiparticles are defined by the complex self-energies and spectral functions. 
The spectral functions can be taken in various forms. Here we consider the most successful physically motivated Lorentzian form \cite{Cassing:2007nb}:
	\begin{align}
		\rho_i(\omega,{\bf p}) = &
		\frac{\gamma_i}{\tilde{E}_{i,\mathbf{p}}} \left(
		\frac{1}{(\omega-\tilde{E}_{i,\mathbf{p}})^2+\gamma_i^2} - \frac{1}{(\omega+\tilde{E}_{i,\mathbf{p}})^2+\gamma_i^2}
		\right)\  \nonumber \\
		& = \frac{4\omega\gamma_i}{\left( \omega^2 - \mathbf{p}^2 - m^2_i \right)^2 + 4\gamma^2_i \omega^2}
		\label{eq:spectral_rho},
	\end{align}
where $i$ runs over the particle species ($i=q, {\bar q}, g$). Here, we introduced the off-shell energy $\tilde{E}_{i,\mathbf{p}} = \sqrt{ {\bf p}^2+m_i^2-\gamma_i^2 }$, $m_i$ and $\gamma_i$ being the particle's pole mass and width.
$\rho_i$ is a real function, odd in $\omega$ and for all ${\bf p}$ it fulfills the sum rule
\begin{equation}
  \int_{-\infty}^{\infty} \frac{d\omega}{2\pi} \omega \rho_i(\omega,{\bf p}) = 1.
\end{equation}

Using the spectral function, the quasiparticle (retarded) propagators can be expressed in the Lehmann representation as:
\begin{equation}
\Delta_i(\omega,{\bf p})=  \int_{-\infty}^{\infty} \frac{d\omega'}{2\pi}\frac{\rho_i(\omega',{\bf p})}{\omega-\omega'}=\frac{1}{\omega^2 -{\bf p}^2 -m_i^2 +2 i\gamma_i\omega }.
\label{eq:prop}
\end{equation}

In \eqref{eq:prop}, we are considering all the (retarded) quasiparticle self-energies to be equal, $\Pi=\Sigma =  \Sigma_q
\approx \Sigma_{\bar q}$, and they are expressed via dynamical masses and widths as: 
\begin{equation}
  \Pi_i = m_i^2 -2i\gamma_i \omega.  
\end{equation}
In the off-shell case, $\omega$ is an independent off-shell energy.
\subsection{DQPM Ansatz}
In the DQPM, the microscopic quantities, i.e. effective masses and thermal widths, depend on the effective coupling constant which, in turn, acquires an explicit temperature and chemical potential dependence ~\cite{Moreau:2019vhw}. This is the main characteristic of the DQPM parametrization used for our first model study.

Now we briefly recall the main details of the parametrization of the DQPM, which has been studied in many variations in Refs \cite{Peshier:2005pp, Cassing:2007nb, Cassing:2007yg, Berrehrah:2016vzw, Moreau:2019vhw, Soloveva:2020hpr, Grishmanovskii:2023gog}:\\
 $\bullet$ In the DQPM, the coupling constant is adjusted by fixing the quasiparticle entropy density to reproduce the entropy density $s(T,\mu_\mathrm{B} = 0)$ from lQCD, e.g.~\cite{Borsanyi:2012cr, Borsanyi:2013bia} (see a parametrization method introduced in Ref. \cite{Berrehrah:2015vhe}).  It has been shown that, for a given value of $g^2$, the ratio $s(T,g^2)/T^3$ is almost constant for different temperatures but identical $g^2$, i.e. ${\frac{\partial}{\partial T}} (s(T,g^2)/T^3)=0$. 
Therefore the entropy density $s$ and the EoS in the DQPM is a function of $g^2$ only, i.e. $s(T,g^2)/s_{SB}(T) = f(g^2)$ where $s^{QCD}_{SB} = 19/9 \pi^2T^3$ corresponds to the Stefan-Boltzmann limit of the entropy density for massless quarks and gluons. Thus, by inverting the $f(g^2)$ function, the coupling constant $g^2$ can be directly obtained from the parametrization of lQCD data for the entropy density $s(T,\mu_B=0)$. The resulting parametrization for the coupling constant reads \cite{Berrehrah:2016vzw}:
	\begin{equation}
	g^2(T,\mu_\mathrm{B} = 0) = d \cdot \Big[ \left(s(T)/s^\mathrm{QCD}_{\mathrm{SB}} \right)^e -1 \Big]^f. 
	\label{coupling_DQPM}
	\end{equation}
Here  the parameters $d = 169.934$, $e = -0.178434$, and $f = 1.14631$ are fixed in accordance with the $s(T)$ calculations by the WB Collaboration from Refs. ~\cite{Borsanyi:2012cr, Borsanyi:2013bia}. \\

$\bullet$  The exact form of pole masses, in the DQPM, depends on the effective coupling constant as predicted by Hard Thermal Loop (HTL) calculations in the high-temperature regime \cite{Bellac:2011kqa, Linnyk:2015rco}.
Considering three flavors and colors, $N_f=N_c=3$, and for vanishing chemical potential, $\mu_q=0$, the gluon ($g$) and light quarks ($u$ and $d$, both labeled with $l$) masses read \cite{Bellac:2011kqa,Linnyk:2015rco}:
\begin{subequations}
    \begin{align}
	   m^2_{g}(T)&=C_a \frac{g^2(T)}{6}T^2\left(1+\frac{N_{f}}{2N_{c}} \right) = \frac{3}{4}g^2(T)T^2 \label{polemassg_dqpm},\\
	  m^2_{l(\bar l)}(T)&= C_f \frac{g^2(T)}{4}T^2 = \frac{1}{3}g^2(T)T^2\label{polemassq_dqpm},
	\end{align}
 \label{eq: mass-dqpm}
 \end{subequations}
where  $C_F = \dfrac{N_c^2 - 1}{2 N_c} = 4/3$ and $C_A = N_c = 3$ are the QCD color factors for quarks and gluons, respectively. It is important to note that here $g^2(T)$ is the effective temperature-dependent coupling constant. \\

$\bullet$ The strange quark mass, whose scaling has not been studied in detail within the HTL framework, is given by the phenomenological equation:
\begin{equation}
    m_{s(\bar{s})}(T)= m_{l(\bar{l})}(T)+ \Delta m_{ls}, \label{eq: strange-mass-dqpm}
\end{equation}
where $\Delta m_{ls}=0.03$ GeV is a constant mass shift. Later on, we will consider how the change of this parameter affect transport coefficients. However, this simple shift is justified by the larger bare mass of the strange quark, which enhances its dynamic mass.
Previously, the value of $\Delta m_{ls}$ has been fixed by comparing experimental data for strange hadrons abundances and the $K^+/\pi^+$ ratio in relativistic heavy-ion collisions using the PHSD, a microscopic covariant transport approach \cite{Moreau:2019vhw}. There, the microscopic properties of the partonic degrees of freedom are described by the DQPM (see latest results and discusions in \cite{Soloveva:2020hpr, Moreau:2019vhw}).\\

$\bullet$ In contrast to the on-shell quasi-particle models \cite{Plumari:2011mk, Li:2022ozl}, where thermal widths are absent and a bag constant must be introduced, in the DQPM the thermal widths are chosen to follow the smooth increase of the interaction rate $\Gamma$ with T. Therefore, the thermal widths have a physical meaning, as they reflect the collision frequency of particles at finite temperature T. 
In the DQPM parametrization, the thermal widths read \cite{Moreau:2019vhw, Berrehrah:2016vzw, Linnyk:2015rco}:
	\begin{equation}
		\gamma_{i}(T) = \frac{1}{3} C_i \frac{g^2(T)T}{8\pi}\ln\left(\frac{2c_m}{g^2(T)}+1\right).
	\label{eq:widths_dqpm}
	\end{equation}
 The constant parameter $c_m = 14.4$ was fixed in Ref.~\cite{Cassing:2007nb} and is related to a magnetic cut-off. The finite width reflects the dynamical modification of the spectral function of quasiparticles during their propagation in the sQGP medium. \\

Furthermore, the thermal widths of all (anti-)quarks are assumed to be equal and completely fixed by $g^2$: $\gamma_{u}=\gamma_{d}=\gamma_{s}$. 
This assumption has been checked in Ref.~\cite{Moreau:2019vhw}, where it was shown that the interaction rates of strange and light quarks coincide, apart from the small difference in the vicinity of the phase transition $T<1.5T_c$.
In the next section, we will discuss possible generalizations of this model using DNNs. 

\subsection{Thermodynamics}\label{sec:thermodynamics}
In this subsection, we detail the precise method by which thermodynamic observables are derived from the microscopic characteristics of off-shell quasiparticles. For the analysis of the main thermodynamic properties within thermal QCD, employing the $\Phi$-functional approach is beneficial. This technique represents the thermodynamic potential $\Omega$ via the dressed propagators $\Delta_i$ \cite{Baym,Blaizot:2000fc}. Following this representation, the derivatives of $\Omega$ can be efficiently calculated. 
In this formalism, the entropy density $s^{\mathrm{dqp}} = - \dfrac{\partial \Omega/V}{\partial T}$ for off-shell quasiparticles can be written as:
\begin{equation}
 s^{dqp} =  2 \int_0^{+\infty} \dfrac{d\omega  dp \ p^2}{4 \pi^3} F_{s}(\omega, p, T, \mu),
\label{sdqp_easy}
\end{equation}
with
\begin{align*}
& F_{s}(\omega, p, T, \mu) = -d_g \frac{\partial f_g ( \omega )}{\partial T} \left( \mathrm{Im}(\ln \Delta^{-1})- \mathrm{Im} \Pi \mathrm{Re}  \Delta \right) \\
& -d_q  \frac{\partial f_q(\omega-\mu_q)}{\partial T} \left( \mathrm{Im}(\ln S_q^{-1})-  \mathrm{Im} \Sigma_q \mathrm{Re}  S_q \right) \\
& -d_{\bar{q}} \frac{\partial f_{\bar q}(\omega+\mu_q)}{\partial T} \left( \mathrm{Im}(\ln S_{\bar{q}}^{-1})- \mathrm{Im} \Sigma_{\bar{q}} \mathrm{Re}  S_{\bar{q}} \right) ,
\end{align*}
where $\Delta_i
=(p^2-\Pi_i)^{-1}$, $S_q = (p^2-\Sigma_q)^{-1}$ and $S_{\bar q} =
(p^2-\Sigma_{\bar q})^{-1}$ stand for the full (scalar) quasiparticle propagator of gluons $g$, quarks $q$, and antiquarks ${\bar q}$. \\
Similarly, the quark density of the off-shell quasi-particles  $n^{\mathrm{dqp}} = -\dfrac{\partial \Omega/V }{\partial \mu}$ reads:
\begin{equation}
 n^{dqp} =  2 \int_0^{+\infty} \dfrac{d\omega  dp \ p^2}{4 \pi^3} F_{n}(\omega, p, T, \mu),
 \label{nbdqp_easy}
\end{equation}
with
\begin{align*}
  &  F_{n}(\omega, p, T, \mu) =  -d_q \frac{\partial f_q(\omega-\mu_q)}{\partial \mu_q} \left( \mathrm{Im}(\ln S_q^{-1})- \mathrm{Im} \Sigma_q \mathrm{Re}  S_q \right) \nonumber \\
  & -d_{\bar{q}} \frac{\partial f_{\bar q}(\omega+\mu_q)}{\partial \mu_q} \left( \mathrm{Im}(\ln S_{\bar{q}}^{-1})- \mathrm{Im} \Sigma_{\bar{q}} \mathrm{Re}  S_{\bar{q}} \right). \nonumber
\end{align*}
In the above formulae, $d_g=2 \times (N_c^2-1)$ is the number of transverse gluonic degrees of freedom while $d_q=d_{\bar{q}}= 2 \times N_c$ is the fermionic one. Furthermore, $f_{g,q}=f_{B,F}$  is the Bose or Fermi distribution for gluon or quark respectively:
\begin{equation}
    f_{B,F}=\left[\exp \left( (E_i - \mu_i)/T\right) \pm 1\right]^{-1},
\end{equation}
where $\mu_{i}$ is the quark chemical potential. The quasi-particle energy is $E_i=\sqrt{\mathbf{p}_i^2+m_i^2}$ for the on-shell case or $\omega_i$ for the considered off-shell case. 

Here we emphasize that the entropy and quark number density 
functionals are not restricted to narrow quasiparticles, i.e. with spectral widths $\gamma_i$ much smaller than the typical energy.

Baryon and strange charge densities follow from the quark densities:
 \begin{align*}
 \nB= n^{\mathrm{dqp}}/3 = 1/3 ( \nuu + \nd + \ns) \\
  \nS = - \ns. 
 \end{align*}
 At vanishing chemical potential, the properties of the quasi-particles are explored using quark-number susceptibilities instead of the densities: 
\begin{equation}
   \chi_i(T,m_i,\gamma_i) = \dfrac{\partial n_i}{\partial \mu_i} \Biggr|_{\mu_i=0}.
\end{equation}

Quark-number susceptibilities are related to the pressure by
\begin{align*}
\frac{\chi_q (T,\mu_q)}{T^2} =\frac{\partial^2 (P/T^4)}{\partial (\mu_q/T)^2},
\end{align*}
and therefore can be related to the  conventional second-order susceptibilities used in the Taylor expansion $\chi_2^{i j}$ at vanishing quark chemical potential as \cite{Borsanyi:2012cr} {\setlength\arraycolsep{0pt}
\begin{eqnarray}
\label{equ:Sec4.7}
\frac{P(T,{\mu_i})}{T^4} = \frac{P(T, {0})}{T^4} + \frac{1}{2} \sum_{i, j} \frac{\mu_i \mu_j}{T^2} \chi_2^{i j}, \nonumber \\
\textrm{with} \ \ \chi_2^{i j} = \frac{1}{T^2} \frac{\partial n_j (T, {\mu_i})}{\partial \mu_i} \Biggr|_{\mu_i = \mu_j = 0}.
\end{eqnarray}}
For the $N_f =3$ case and transverse gluons, the total entropy and the baryon and strange susceptibilities read:
\begin{subequations}\label{thermodynamic functions DQPM}
\begin{align}
    s(T)&=-d_g I^B_g(T) -d_q\sum_{i=q,s} I^F_i(T) ,\label{entropy}\\
    \chi_2^B(T) &= \frac{1}{T^2}\frac{d_q}{9}(2\chi_2^{l}(T) + \chi_2^{s}(T)),\label{chib}\\
    \chi_2^S(T) &= \frac{1}{T^2} d_q\chi_2^{s}(T),\label{chis}
\end{align}
\end{subequations}
where we have introduced the integrals:
\begin{widetext}    
\begin{subequations}\label{I and chi integrals}
    \begin{align}
        I^{B,F}_{i}(T,m,\gamma) =& \frac{1}{2\pi^2 T}\int \di \, p^2\frac{4p^2+3m_i^2}{3\sqrt{\omega^2+p^2}}f_{B,F}(\omega,T)+2\int_0^\infty \frac{\di\omega}{2\pi}\int\frac{\di^3p}{(2\pi)^3}\frac{\partial f_{B,F}(\omega,T)}{\partial T}h(\omega,p,m_i,\gamma_i),\label{IBF integral}\\
        \chi^i_2(T,m,\gamma) =&\frac{1}{2\pi^2 T}\int_0^{\infty}\di p\,
        \frac{p^2}{1+\cosh\left(\sqrt{m_i^2+p^2}/T\right)}+ 2\int_0^{\infty}\frac{\di\omega}{2\pi}\int \frac{\di^3p}{(2\pi)^3}\frac{\sinh(\omega/T)}{T^2(1+\cosh(\omega/T))^2}h(\omega,p,m_i,\gamma_i),\label{chi_integral}
    \end{align}
\end{subequations}
where the function $h$ is the auxiliary function:
\begin{equation}
    h(\omega,p,m_i,\gamma_i)=2\gamma_i\omega\frac{\omega^2-\bm{p}^2-m_i^2}{(\omega^2-\bm{p}^2-m_i^2)^2+4\gamma_i^2\omega^2}-\arctan\left(\frac{2\gamma_i\omega}{\omega^2-\bm{p}^2-m_i^2}\right).
\end{equation}
\end{widetext}
Notice that all thermodynamical functions implicitly depend on $g^2$ through the dressed masses and widths if we consider a specific parametric ansatz as in the DQPM.
An alternative approach involves a traditional perturbative series expansion of the entropy with respect to the coupling constant $g$ as in references \cite{Kajantie:2002wa, Kraemmer:2003gd}:
\begin{equation}
s/T^3 = c_0 + c_2g^2 + c_3g^3 + \dots
\end{equation}
Here, $c_0 = \frac{\pi^2}{45}(4(N_c^2 - 1) + 7 N_c N_f)$ corresponds to the Stefan-Boltzmann limit. Moreover, there are different resummations techniques for thermal QCD such as hard-thermal-loop perturbation theory (HTLpt), which are not discussed here (for further reading we refer to Ref. \cite{Mogliacci:2013mca}). To enhance the convergence of the HTL resummation, a self-consistent quasi-particle expansion has been adopted, as described in Ref. \cite{Blaizot:2000fc}.

From the entropy density and quark densities, other thermodynamic quantities follow.
The pressure at vanishing baryon chemical potential $\mu_B=0$ is defined employing the entropy density as
\begin{align}
	p(T) =  p^{lqcd}(T_0)  + \int\limits_{T_{0}}^{T} s(T')\ dT'  \label{pressure0} ,
\end{align}
where $p^{lqcd}(T_0)$ is taken from lQCD after fixing $T_0$.   

The energy density $\epsilon$ then follows from the Euler relation
\begin{equation}
	\label{eps}  \epsilon = T s - p.% + \sum_i \mu_i n_\mathrm{i}, where in general $i=B,Q,S$.
\end{equation}

Another important thermodynamic observable is the trace of the energy-momentum tensor, also known as interaction measure or trace anomaly:
\begin{equation}
    I = \epsilon - 3 p = T s - 4p.
\label{trace_anom}
\end{equation}
From lQCD calculations \cite{Borsanyi:2011sw, Borsanyi:2013bia, Borsanyi:2022qlh, HotQCD:2014kol}, the trace anomaly $I$ is expected to be sizeable in the vicinity of the cross-over transition, indicating a strong interaction of the medium. Consequently, one can expect the shear viscosity close to $T_c$ to be correspondingly smaller than for a weakly interacting medium, as considered in the pQCD limit.

\subsection{Transport coefficients}
As complementary observables, transport coefficients can reveal how well the considered models can microscopically describe the  dense QGP medium.
Therefore, we aim to evaluate transport coefficients using the Kubo-Zubarev formalism, where we employ the parton spectral functions without relying on the relaxation time approximation or effective coupling constant. Since quasiparticles masses and widths - obtained from the DNNs - can be larger than those in the original DQPM, this method is better suited.

The shear viscosity is evaluated from the slope of the Fourier transform of the spectral function for the spatial traceless part of the stress tensor $\langle[\pi_{ij}(x),\pi_{ij}(0)]\rangle$ in the limit $\omega \rightarrow 0$. 
Here we employ the following formula for the shear viscosity:
\begin{align}
\eta^{\text{Kubo}}(T) & = - \int \frac{d^4p}{(2\pi)^4}\ \sum_{i=q,\bar{q},g} d_i\ \frac{\partial f_i(\omega)}{\partial \omega}\ \rho_i(\omega,\mathbf{p})^2 \Pi_{i} \nonumber\\
=  \frac{1}{15T} \int & \frac{d^4p}{(2\pi)^4}\ \sum_{i=q,\bar{q},g} d_i \left[ (1 \pm f_i(\omega)) f_i(\omega) \right] \rho_i(\omega,\mathbf{p})^2 \Pi_{i} , 
\label{eta_Kubo}
\end{align}
where the notation $f_i(\omega) = f_i(\omega,T,\mu_q)= f_{B,F}$ is used for the distribution functions. The corresponding derivative of the distribution function accounts for the Pauli-blocking ($-$) and Bose-enhancement ($+$) factors, and $\rho_i$ denotes the spectral functions from Eq. (\ref{eq:spectral_rho}). 
Using the notation $\Pi_{i=q,g}$ we differentiate the contribution from transverse gluons \cite{Aarts:2002cc, Peshier:2005pp}:
\begin{equation}
     \Pi_{g} = 7 \omega^4  -10 (\omega\mathbf{p})^2+ 7\mathbf{p}^4,  
\end{equation}
and from quarks \cite{Aarts:2002cc, Lang:2012tt}:
\begin{equation}
    \Pi_{q} =  p_x^2 p_y^2.
\end{equation}

We note that (for weak coupling) it is common to derive $\eta$ from a Boltzmann equation to next-to-leading log (NLL) order \cite{Arnold:2003zc}:
 \begin{equation}\label{eq: eta NLL}
     \eta^{NLL} \approx \dfrac{T^3}{g^2 \ln (1/g) }.
 \end{equation}
 This approach is suited for on-shell or narrow quasi-particles, whereas we don't assume small $g$ and employ the more rigorous Kubo-formalism. It is expected that, near $T_c  \approx 158 $ MeV, the results of Eq.  \eqref{eq: eta NLL} are not applicable.
 
%%%%%%%%%%%%%%%conductivity
We consider here another important transport coefficient, i.e. the electric conductivity for stationary electric fields  $\sigma_Q(T)$, which describes the response of the system to an external electric field.
In the case of the electrical conductivity only quark degrees of freedom contribute, resulting in an analogous expression \cite{Harutyunyan:2017ttz}:
\begin{align}
\sigma_Q^{\text{Kubo}}(T) & = - \int \frac{d^4p}{(2\pi)^2}\  \sum_{i=q,\bar{q}} d_i\ \frac{\partial f_i(\omega)}{\partial \omega} \rho_i(\omega,\mathbf{p})^2  \nonumber\\
=  \frac{1}{3T} \int & \frac{d^4p}{(2\pi)^4}\ \mathbf{p}^2 \sum_{i=q,\bar{q}} d_i \left[ (1 \pm f_i(\omega)) f_i(\omega) \right] \rho_i(\omega,\mathbf{p})^2  , 
\label{eq: sigma_Kubo}
\end{align}
~\\
for $q=u,d,s$.
By separating the contributions of the strange flavor one can also single out the strange quark's contribution to the conductivity, which gives us a complementary observable to infer the properties of the strange quarks.

%------------------------------------------------------------------
\section{\label{sec-dnn-scheme} Neural networks for the regression applied to the quasi-particle model}
In this section, we describe the technical details of our framework, covering aspects such as the input and output observables, the architecture of the DNNs, the training process, and the evaluation methods.
The code of the neural network is written in Python and it is implemented using the Keras Deep Learning API v2.13.1 \cite{chollet2015keras} together with the Tensorflow v2.13.0 library \cite{Abadi:2016kic}.

For the regression task, we utilize Feed-forward neural networks (FFNNs), which are a preferred choice over CNN especially for simple, non-spatially-distributed datasets, due to their efficiency and the smaller number of hyperparameters, allowing a faster training. The number of layers of the DNNs is chosen heuristically, balancing between the quality of the fit, the smaller number of parameters, and the training speed. Three hidden layers, with 24, 12, and 12 neurons respectively, turn out to be sufficient for the purpose of this study. For all layers, the activation function is \emph{sigmoid}:
\begin{equation}
    \sigma(x)=\frac{1}{e^x+1}.
\end{equation}

For the minimization procedure, we employ the Adam algorithm \cite{Kingma:2014vow}.
For all of the presented results, the learning rate is initially set to 0.005 and decays by 90\% every 1000 epochs of the training. The number of epochs used in the training is $4\times 10^3$, and we observed that increasing this number by a factor of three doesn't change our results and leads to overfitting.  

In this study, we use machine learning to infer the functional form of microscopic quasiparticles properties (masses, widths and coupling constant) as a function of temperature, such that the formulae obtained with the $\Phi$-functional approach - detailed in Sec.~\ref{sec:thermodynamics} - are in agreement with lattice QCD data. Therefore, we consider a DNN-based DQPM model, which we will refer to as DQPMnn. The input of the DQPMnn is the temperature, expressed in GeV, in the range $T>T_c$, $T_c = 0.158$  GeV in accordance with the employed lQCD data from the WB collaboration. 

We will explore two distinct regression tasks. As outputs for these tasks, we consider (i) the squared coupling constant $g^2(T)$; and (ii) a combination of $g^2(T)$, the masses $m_i(T)$, and the widths $\gamma_i(T)$ of the quasi-particles. The next subsections describe the details of the loss function, and the extraction of the physical properties of the quasiparticles.

%%%%%%%%%%%%%transport coefficients
 \subsection{DQPMnn: $\mathcal{L}_{0}$ - extraction of $g^2(T)$}\label{sec:DQPM g2}
The original idea of DQPM relies on adjusting the effective coupling to fit the entropy density, employing an HTL-based parametrization for the effective masses and widths. In this section, as proof of principle, we employ the DQPM parametrization, Eqs. \eqref{eq: mass-dqpm}, \eqref{eq: strange-mass-dqpm} and \eqref{eq:widths_dqpm}, and compare the outputs of the DQPMnn to different lattice observables, such as the EoS and the transport coefficients. In what follows, and throughout the paper, in order to distinguish quantities from the the DQPMnn, we use an underline to identify them. For example, $\underline{g}^2(T)$ is the function associated with the running coupling output by the DQPMnn, whereas $g^2(T)$ is the one used in the original DQPM.

To extract the coupling constant $\underline{g}^2(T)$ with the DQPMnn, we use a loss function:
 \begin{align}
 \mathcal{L}_0= \beta_{G} \left[\frac{s(T)/T^3-s_\text{lQCD}/T^3}{\Delta s_{lQCD}/T^3}\right]^2\nonumber \\
 + \beta_{L}  \left[\frac{\chi_2^B(T)-{\chi^B_2}_\text{lQCD}}{\Delta {\chi^B_2}_{lQCD}}\right]^2 \nonumber \\
 + \beta_{S} \left[\frac{\chi_2^S(T)-{\chi^S_2}_\text{lQCD}}{\Delta {\chi^S_2}_{lQCD}}\right]^2 .
 \label{eq:loss dqpm}
\end{align}
The minimization of the above loss function is analogous to the minimization of the $\chi^2$ in the standard fitting procedures, but the parameters 
%$\beta_{G,L,S}$ 
$\beta_G, \beta_L, \beta_S$
allow to regulate the contribution of each thermodynamic quantity to the loss function. Notice that the loss function in eq.~\eqref{eq:loss dqpm} is not only a minimum squared error, as it was used for example in \cite{Li:2022ozl}, but also takes into account the uncertainties associated with the lattice measurements.

The case $\beta_G = 1$ and $\beta_L=\beta_S=0$, where microscopic quantities are inferred from the entropy density only, should reproduce the results of the DQPM. The thermodynamic functions are computed from Eqs. \eqref{thermodynamic functions DQPM} and the subscript ``lQCD'' labels the lattice data, which have been taken from \cite{Borsanyi:2011sw, Borsanyi:2013bia, Borsanyi:2022qlh}. All the thermodynamic functions appearing in the loss have an implicit dependence on $\underline{g}(T)$ through the mass and width of the different quasi-particles. To be more explicit, one should write, for example, $s(T)\rightarrow s(T,\underline{g}(T)$), but we choose to simplify the notation and drop the implicit dependence of the thermodynamic functions.

Notice that, to train the DQPMnn, we have chosen thermodynamic quantities that are dimensionless.  The main advantage of the use of dimensionless observables such as $s/T^3$, $\chi_2^B$, $\chi_2^S$ is that they mitigate the impact of discrepancies between lattice EoS results obtained by different groups: HotQCD \cite{HotQCD:2014kol} or WB Collaboration \cite{Borsanyi:2013bia, Borsanyi:2012cr,Borsanyi:2022qlh} results influence less the final results for the desired microscopic quantities: masses, widths.  In contrast, if we consider $s(T)$ in $\text{GeV}^3$ as an input for the training,  the difference between the two EoS would be larger. 
This is in contrast with a recent work, Ref.~\cite{Li:2022ozl}, where DNNs were used for an on-shell quasi-particle model with a bag constant and $s(T)$ and $\Delta=I (T)$ were used for training. 

However, the loss function in Eq. \eqref{eq:loss dqpm} requires the evaluation of thermodynamic integrals at each step during the training process. While the computation of one of the integrals in Eqs. \eqref{I and chi integrals} is rather fast, the frequency at which they must be calculated renders the use of Eq. \eqref{eq:loss dqpm} not optimal, as it would significantly prolong the training duration.
To overcome this complication, we use other 2 neural networks as \emph{surrogate models} to approximate the integrals in Eqs. \eqref{I and chi integrals}, and hence $s(T)$ and $\chi_2^{B,S}(T)$. These neural networks take $T$, $m/T$ and $\gamma/T$ as input and are trained to reproduce the values of the integrals $I^{B,F}$ and $\chi^i_2(i = l,s)$  in Eqs. \eqref{I and chi integrals}. A schematic representation of the surrogate models is given in Fig. \ref{fig:scheme_surr}.  Denoting the functions resulting from the surrogate model with a tilde ($\tilde{I}^{F,B}$ and $\tilde{\chi_2}$) to distinguish them from the actual values of Eqs. \eqref{I and chi integrals}, we can use a loss function with the substitutions $s\rightarrow \tilde{s}$ and $\chi_2^{B,S}\rightarrow \tilde{\chi}_2^{B,S}$, where the tilded thermodynamic functions are computed from Eqs. \eqref{thermodynamic functions DQPM} using $\tilde{I}^{F,B}$ and $\tilde{\chi}_2$ in place of $I^{F,B}$ and $\chi_2$:
\begin{subequations}\label{thermodynamic functions DQPM with tildes}
\begin{align}
    \tilde{s}(T)&=-d_g \tilde{I}^B_g(T) -d_q\sum_{i=q,s} \tilde{I}^F_i(T) ,\label{entropy with tilde}\\
    \tilde{\chi}_2^B(T) &= \frac{1}{T^2}\frac{d_q}{9}(2\tilde{\chi}_2^{l}(T) + \tilde{\chi}_2^{s}(T)),\label{chib with tilde}\\
    \tilde{\chi}_2^S(T) &= \frac{1}{T^2} d_q\tilde{\chi}_2^{s}(T).\label{chis with tilde}
\end{align}
\end{subequations}
The use of the surrogate model leads to a significant improvement in the speed of the training. 
The actual loss function used in the training therefore becomes: 
\begin{align}
 \mathcal{L}_0= \beta_{G} \left[\frac{\tilde{s}(T)/T^3-s_\text{lQCD}/T^3}{\Delta s_{lQCD}/T^3}\right]^2\nonumber \\
 + \beta_{L}  \left[\frac{\tilde{\chi}_2^B(T)-{\chi^B_2}_\text{lQCD}}{\Delta {\chi^B_2}_{lQCD}}\right]^2 \nonumber \\
 + \beta_{S} \left[\frac{\tilde{\chi}_2^S(T)-{\chi^S_2}_\text{lQCD}}{\Delta {\chi_2^S}_{lQCD}}\right]^2 .
 \label{eq:loss-0-dqpm}
\end{align}
Further details about the surrogate model are reported in Appendix \ref{appex:surrogate}.

\begin{figure}[h]
    \centering
    \includegraphics[width=0.4\textwidth]{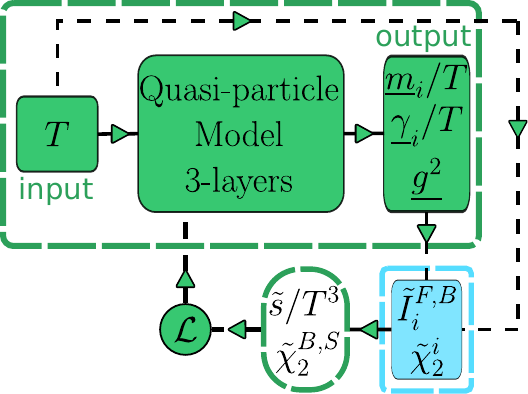}
    \caption{The representation of the quasiparticle DNN model, DQPMnn. An input temperature is used to generate as output the expected coupling constant $g$, as well as two key parameters: the dimensionless masses $m_i/T$ and widths $\gamma_i/T$ for each particle species. The latter two parameters are then fed into the surrogate model, in light blue in the figure, to obtain the functions $\tilde{I}$ and $\tilde{\chi_2}$. These functions are used to compute the entropy density and the susceptibilities as described in the text. The loss function minimizes the squared error between the neural network's guess and the lattice data.}
    \label{fig:schemenn}
\end{figure}

\begin{figure}[h]
    \centering
    \includegraphics[width=0.4\textwidth]{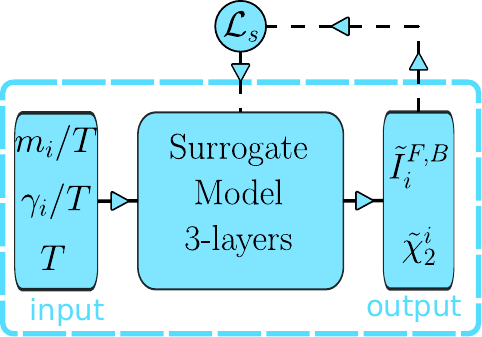}
    \caption{A schematic representation of the neural network used for the surrogate model. The input data, $T$, $m/T$, and $\gamma/T$ are used to evaluate the functions $\tilde{I}$ and $\tilde{\chi}_2$ in Eqs. \eqref{I and chi integrals}. The training uses the mean squared error as a loss function, such that $\tilde{I}^{F,B}$ and $\tilde{\chi}_2$ reproduce $I^{F,B}$ and $\chi_2^i$ evaluated numerically from Eqs. \eqref{I and chi integrals}. The complete procedure is described in Appendix \ref{appex:surrogate}.}
    \label{fig:scheme_surr}
\end{figure}

The structure of the DQPMnn is depicted schematically in Figure \ref{fig:schemenn}. However, in this section, we are only considering $\underline{g}$ as an output of the DQPMnn, whereas the masses and the width are taken from the DQPM parametrization, Eqs. \eqref{eq: mass-dqpm}, \eqref{eq: strange-mass-dqpm} and \eqref{eq:widths_dqpm}, with the substitution $g\rightarrow\underline{g}$. The consequences of relaxing the DQPM parametrization will be explored in the forthcoming sections.  \\

%%%%%%%%%%%%%%%%%%%%%%%%%%%%%%%%%% 
%main scheme
%%%%%%%%%%%%%%%%%%%%%%%%%%%%%%%%%% 

%%%%%%%%%%%%%%%%%%%%%%%%%%%%%%%%%%%%%%%%%%%%%%%%%%%%%%%
% surrogate scheme
%%%%%%%%%%%%%%%%%%%%%%%%%%%%%%%%%%%%%%%%%%%%%%%%%%%%%%%

 We have examined different values of the weights 
$\beta_G, \beta_L, \beta_S$
 in the loss function $\mathcal{L}_0$ to understand how these adjustments influence the microscopic quantities, such as masses, widths, and the coupling constant, based on the selected thermodynamic observable.
 In particular, we have tested the three setups resulting from setting one of the $\beta_i$ to one and the others to zero. 
 These setups called ``I'', ``II'' and ``III'', are explicitly summarized in Table \ref{table:weights L0}.
 They correspond to 
 \begin{itemize}
\item  setup ``I'' -- best DNN fit of $\mathcal{L}_0$ to $s/T^3$ (notations in the figures ``$\mathcal{L}_0: s/T^3$''),
\item setup ``II'' -- best DNN fit of $\mathcal{L}_0$ to $\chi_2^B$ (notations in the figures ``$\mathcal{L}_0: \chi_2^B$''),
\item setup ``III'' -- best DNN fit of $\mathcal{L}_0$ to $\chi_2^S$ (notations in the figures``$\mathcal{L}_0: \chi_2^S$'').
 \end{itemize}  

 \begin{table}[h!]
     \centering
     \begin{tabular}{|c|c|}
     \hline
          setups & $\beta_G$ : $\beta_L$ : $\beta_S$   \\
    \hline
          I      &  $1$ : $0$ : $0$\\
    \hline
          II      &  $0$ : $1$ : $0$\\
    \hline
          III      &  $0$ : $0$ : $1$\\
    \hline
     \end{tabular}
     \caption{The three setups used to test the DQPM model parametrization on different observables. The weights refer to the loss $\mathcal{L}_0$ in Eq. \eqref{eq:loss-0-dqpm}. }
     \label{table:weights L0}
 \end{table}
 %%%%%%%%%%%%%%%%%%%%%%%%%%%%%%%%%%%%%%%%%%%%%%%%%%%%%%%%
 
We begin with the comparison of the thermodynamic observables used for the minimization of the loss function $\mathcal{L}_0$ to the true values - the lQCD estimates.
Fig. \ref{fig:eos_dqpm_ansatz} shows the dimensionless entropy ($s/T^3$) in the top panel, baryon susceptibility ($\chi_2^B$) in the middle panel, and strangeness susceptibility ($\chi_2^S$) in the bottom panel, all as functions of the scaled temperature ($T/T_c$). The lines represent predictions from the DQPMnn with $\mathcal{L}_0$ in setups I (red solid lines), II (blue solid lines), and III (green solid lines). The symbols correspond to the true values - estimates from lattice QCD by the WB Collaboration \cite{Borsanyi:2011sw, Borsanyi:2013bia, Borsanyi:2022qlh}. We observe that the setup I underestimates the susceptibility, similar to the DQPM model. Setups II and III are very similar to each other and overestimate the entropy.

\subsection{Results: $g^2, m, \gamma$ and transport coefficients.}
Now, let's delve into the microscopic characteristics, starting with the output of DQPMnn - $g^2$. For a comparison with the previous results, the running coupling $\alpha_S = g^2/(4 \pi)$ is displayed in Fig \ref{fig: alphas-loss0} as a function of the scaled temperature $T/T_c$ in the setups I-III and is compared to the DQPM result.
%%%%%%%%%%%%%%%%%% eos
\begin{figure}
\centering
\begin{minipage}[h]{0.95\linewidth}
\center{\includegraphics[width=0.99\linewidth]{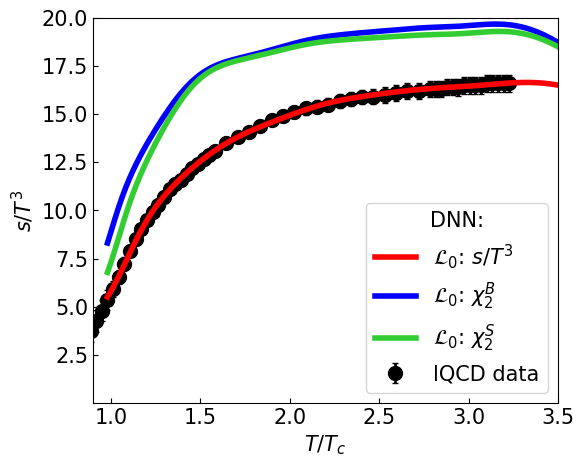}}
\end{minipage}
\begin{minipage}[h]{0.95\linewidth}
\center{\includegraphics[width=0.99\linewidth]{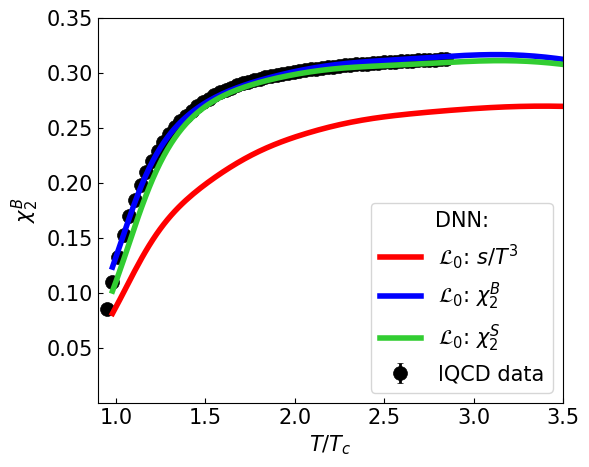}}
\end{minipage}

\begin{minipage}[h]{0.95\linewidth}
\center{\includegraphics[width=0.99\linewidth]{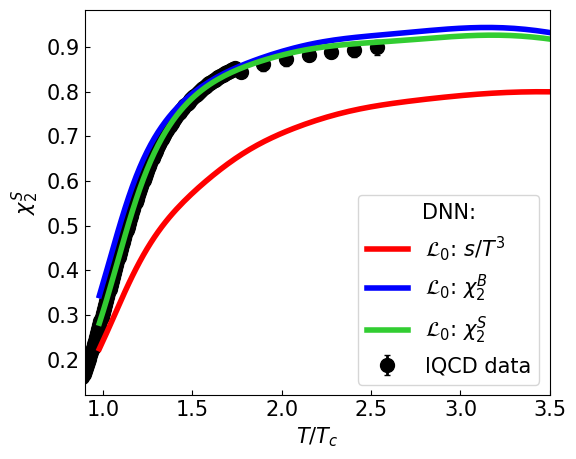}}
\end{minipage}
\caption{The dimensionless entropy $s/T^3$ (top panel), baryon susceptibility $\chi_2^B$ (middle panel) and strangeness susceptibility  $\chi_2^S$ (bottom panel) as a function of the scaled temperature $T/T_c$.  Lines correspond to the predictions of the DQPMnn with $\mathcal{L}_0$ in setups I (red solid lines), II (blue solid lines), and III (green solid lines). The symbols correspond to the true values - lQCD results from the WB Collaboration \cite{Borsanyi:2011sw, Borsanyi:2013bia, Borsanyi:2022qlh}. }
\label{fig:eos_dqpm_ansatz}
\end{figure}
%%%%%%%%%%%%%%%%%%%%%%%%%%%%%%%%%

%%%%%%%%%%%%%%% g2,masses, widths
\begin{figure}[h!]
\centering
\includegraphics[width=0.5\textwidth]{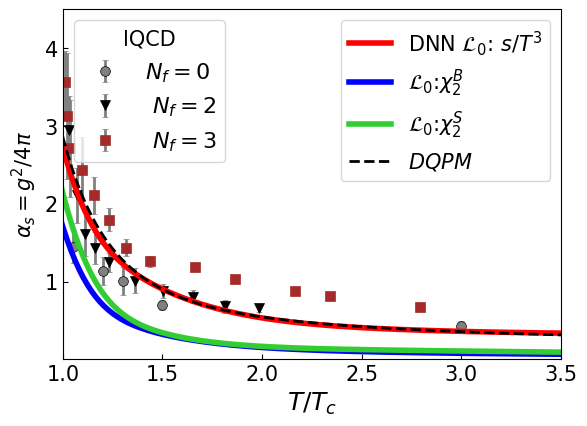}
\caption{The running coupling $\alpha_S = g^2/(4 \pi)$ as a function of the scaled temperature $T/T_c$ constrained from the DQPMnn with $\mathcal{L}_0$  in the setups I-III is compared to the DQPM parametrization (black dashed lines). The color coding of the different setups is the same as in Fig. \ref{fig:eos_dqpm_ansatz}. The lattice
results for quenched QCD, $N_f = 0$, (grey circles) are taken from Ref. \cite{Kaczmarek:2004gv}, $N_f = 2$ (black triangles)  from Ref. \cite{Kaczmarek:2005ui}, and for $N_f =2 +1$ (brown square) from Ref. \cite{Kaczmarek:2007pb}.}
\label{fig: alphas-loss0}
\end{figure}
%%%%%%%%%%%%%%%%%%%%%%%%%%%%%%%%%

\begin{figure}[!h]
\centering
\begin{minipage}[h]{0.9\linewidth}
\includegraphics[width=0.98\linewidth]{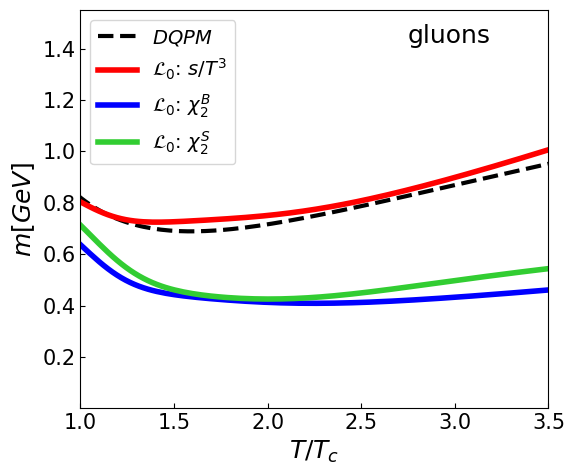}% \\ b) 
\end{minipage}
 \begin{minipage}[h]{0.9\linewidth}
\includegraphics[width=0.98\linewidth]{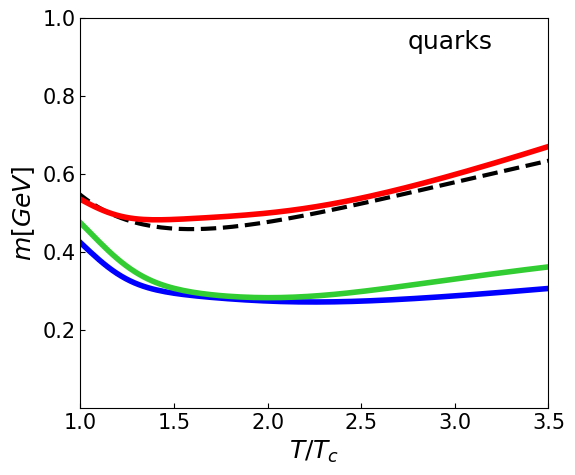}% \\ b) 
 \end{minipage}
 \caption{Masses [GeV] as a function of the scaled temperature $T/T_c$ for gluons (top panel) and light quarks (bottom panel) from the DQPMnn with the loss function $\mathcal{L}_0$ in the setups I-III in comparison to the standard DQPM parametrization (black dashed lines). The color coding for different setups is the same as in Fig. \ref{fig:eos_dqpm_ansatz}.}
  \label{fig:dqpm_mass_evolution}
    \end{figure}

\begin{figure}[!h]
\centering
\begin{minipage}[h]{0.92\linewidth}
\includegraphics[width=0.98\linewidth]{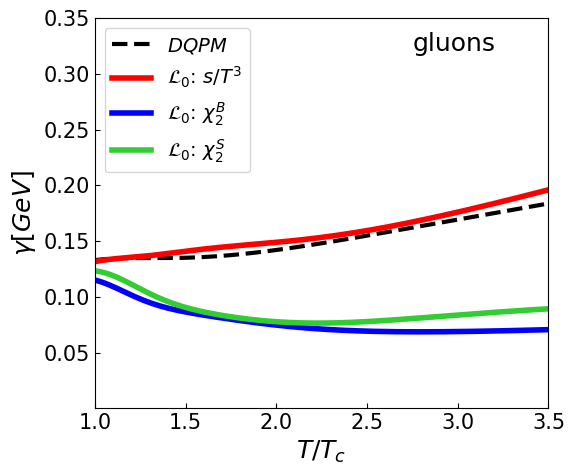} 
\end{minipage}
 \begin{minipage}[h]{0.92\linewidth}
\includegraphics[width=0.98\linewidth]{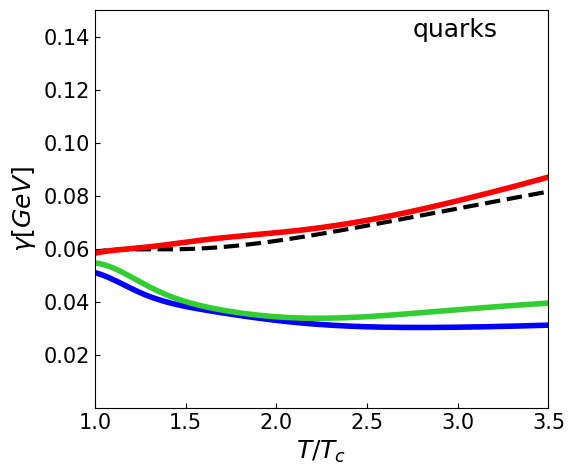}% \\ b) 
 \end{minipage}
 \caption{Widths [GeV] as a function of the scaled temperature $T/T_c$ for gluons (top panel) and light quarks (bottom panel) from the DQPMnn with loss function $\mathcal{L}_0$ in the setups I-III in comparison to the standard DQPM parametrization (black dash-dotted line). The color coding for different setups is the same as in Fig. \ref{fig:eos_dqpm_ansatz}.}
  \label{fig:dqpm_wi_evolution}
    \end{figure}

%%%%%%%%%%%%%---transport coefficients
%%%%%%%%%%%%%-etas
    \begin{figure}[!h]
 \centering
\begin{minipage}[h]{0.985\linewidth}
\includegraphics[width=0.98\linewidth]{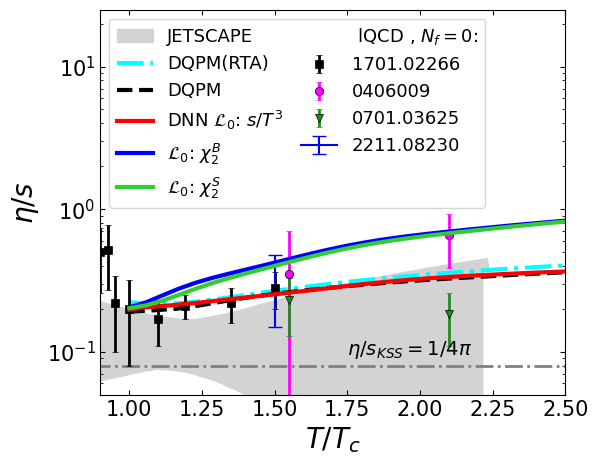}% \\ b) 
\end{minipage}
\caption{Specific shear viscosity  $\eta/s$  as a function of the scaled temperature $T/T_c$ constrained from the DNN with the loss function $\mathcal{L}_0$ in the setups I-III estimated from Eq.~\eqref{eta_Kubo}. The color coding for different setups is the same as in Fig. \ref{fig:eos_dqpm_ansatz}. The black and cyan dashed lines correspond to the DQPM estimates for Kubo (Eq. \eqref{eta_Kubo}) \cite{Moreau:2019vhw} and RTA with the interaction rate formalism \cite{Soloveva:2020hpr}, respectively. 
The dashed gray line demonstrates the Kovtun-Son-Starinets bound \cite{Policastro:2001yc,Kovtun:2004de} $(\eta/s)_{\text{KSS}} = 1/4\pi$.  The symbols show lQCD data for pure SU(3) gauge theory taken from Refs. \cite{Astrakhantsev:2017nrs} (black squares), \cite{Altenkort:2022yhb} (blue line),
\cite{Nakamura:2004sy} (green triangles), \cite{Meyer:2007ic} (magenta circles). The grey area corresponds to the latest estimates from a Bayesian analysis by the JETSCAPE Collaboration \cite{JETSCAPE:2020shq}.}
  \label{fig:etas-sigmas-loss-0}
    \end{figure}

%%%%%%%%%%%%%-sigmas
\begin{figure}[!h]
\centering
\begin{minipage}[h]{0.995\linewidth}
\includegraphics[width=0.98\linewidth]{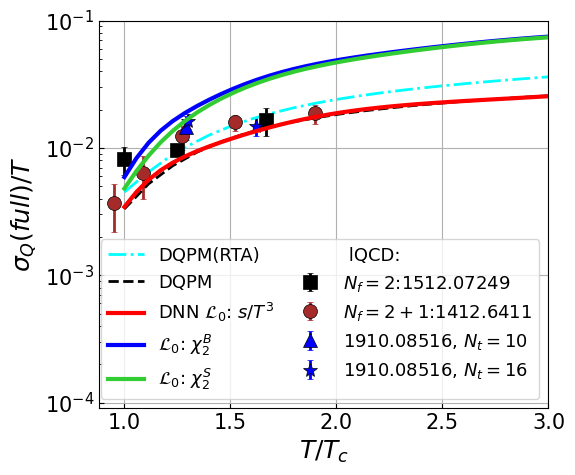}% \\ b) 
\end{minipage}
 \begin{minipage}[h]{0.995\linewidth}
\includegraphics[width=0.98\linewidth]{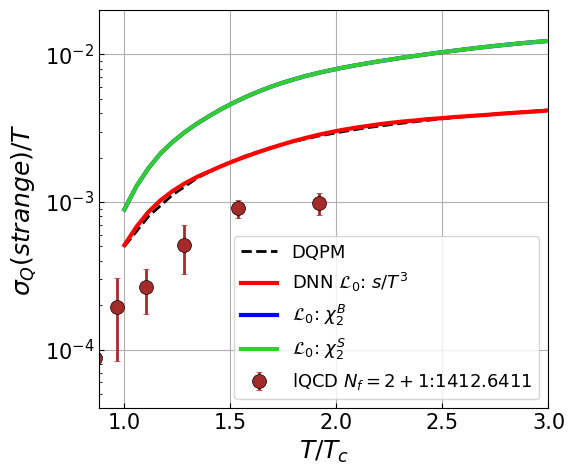}% \\ b) 
 \end{minipage}
\caption{Scaled electric conductivity $\sigma_Q/T$ for all flavors (top panel) and strange quarks (bottom panel) as a function of the scaled temperature $T/T_c$ constrained from the DQPMnn with the loss function $\mathcal{L}_0$ in the setups I-III estimated from Eq. (\ref{eq: sigma_Kubo}). The color coding for different setups is the same as in Fig. \ref{fig:eos_dqpm_ansatz}. Black solid and cyan lines correspond to the DQPM estimates for Kubo (Eq. (\ref{eq: sigma_Kubo})) and RTA with the interaction rate formalism \cite{Soloveva:2020hpr}, respectively. The symbols correspond to the lQCD results. In the top panel, black squares represent $N_f=2$ data from Ref.~\cite{Brandt:2015aqk}, whereas $N_f=2+1$ data are blue triangles and stars from Ref.~\cite{Astrakhantsev:2019zkr}, and brown circles from Ref.~\cite{Aarts:2014nba}. The latter reference is also used for the strangeness contribution to the conductivity in the bottom panel, where lattice data are brown circles.}
  \label{fig:sigmas-loss-0}
    \end{figure}
Additionally, we show lQCD estimates:  quenched QCD $N_f = 0$ results (grey circles) are taken from Ref. \cite{Kaczmarek:2004gv},  $N_f = 2$ (black inverted triangles) from Ref. \cite{Kaczmarek:2005ui} and $N_f = 2+1$ (brown squares) from Ref. \cite{Kaczmarek:2007pb} (from Fig. 7). It is important to note that the coupling constant obtained by lQCD strongly depends on the definition of $\alpha_s$ extracted from the static potential \cite{Kaczmarek:2004gv, Kaczmarek:2007pb}. The main feature of the resulting coupling constants is a significant increase approaching $T_c$, comparable to the predictions from lQCD, showing the importance of nonperturbative effects for $T< 3T_c$. 

One can see that the initial choice of fitting $s/T^3$ in the case of setup I aligns with the DQPM parametrization.  On the other hand, setups II and III lead to an unrealistically low coupling.

In Figs. \ref{fig:dqpm_mass_evolution}, \ref{fig:dqpm_wi_evolution} we show the masses and widths for gluons (top panels) and light quarks (bottom panels) as a function of $T/T_c$ for different setups and compare them to the benchmark DQPM results (black dash-dotted line).
As seen from the figures, in the DQPM and DQPMnn $\mathcal{L}_{0}$ setup I, the masses and widths of quarks and gluons increase with $T$. For all presented setups the results respect the colour factors $M_q = \frac{2}{3} M_g, \gamma_q = \frac{4}{9} \gamma_g$.
The temperature dependence of the masses and widths are rather different for setups II and III since they follow the extracted $g^2$. Similarly the $m_i, \gamma_i$ for setups II and III are smaller than the DQPM one. Notable small differences are observed between setups II and III, particularly near the critical temperature $T_c$ and for temperatures exceeding $2.5 T_c$. This observation aligns well with our initial expectations, which are based on the comparative analysis of thermodynamic observables used for training and prediction, as depicted in Fig. \ref{fig:eos_dqpm_ansatz}.

 It's important to highlight that fitting all three thermodynamic observables using the basic DQPM Ansatz is not feasible: as we have just shown fitting the entropy density underestimates the susceptibilities, whereas fitting the latter overestimates the entropy density. Thus, in the subsequent section, we will modify the DQPM parametrization to potentially extend the model for a more comprehensive representation of all thermodynamic observables.
 
 Similarly to the thermodynamic observables, we compare the transport coefficients from the Kubo-Zubarev method for three setups to the estimates of the original DQPM spectral functions. \\

 To ensure an accurate comparison we focus on dimensionless transport coefficients. These include the shear viscosity to entropy density ratio, denoted as $\eta/s$, and the ratio of electric conductivity to temperature, expressed as $\sigma_Q/T$.
 Fig. \ref{fig:etas-sigmas-loss-0} displays the specific shear viscosity $\eta/s$ as a function of the scaled temperature $T/T_c$ in comparison to various results from the literature. Multiple colored lines represent the DQPMnn estimates from Eq. (\ref{eta_Kubo}) using three different setups for the weights. The black and cyan dashed lines correspond to the DQPM estimates for Kubo (Eq. (\ref{eta_Kubo})) $\eta^{\text{Kubo}}/s$ \cite{Moreau:2019vhw} and RTA with the interaction rate $\eta^{\text{RTA}}/s$ \cite{Soloveva:2020hpr}, respectively.
 The RTA estimate of the shear viscosity is found to be very close to the one from the Kubo formalism  \cite{Moreau:2019vhw} indicating that the quasiparticle limit ($\gamma \ll M$) holds in the DQPM Ansatz.
 
We also present the primary estimates obtained from well-established methods such as lQCD calculations, AdS/CFT correspondence, and Bayesian analysis, which can serve as valuable guidance for further improvements, although these estimates still rely on certain assumptions about the QGP.
A dashed gray line highlights the well-known Kovtun-Son-Starinets bound $(\eta/s)_{\text{KSS}} = 1/(4\pi)$ \cite{Policastro:2001yc,Kovtun:2004de}. 
Symbols on the plot represent lQCD data for pure SU(3) gauge theory from various references: black squares from \cite{Astrakhantsev:2017nrs}, a blue line from \cite{Altenkort:2022yhb}, green triangles from \cite{Nakamura:2004sy}, and magenta circles from \cite{Meyer:2007ic}. The grey area encompasses estimates from the Bayesian analysis conducted by the JETSCAPE Collaboration \cite{JETSCAPE:2020shq}. We see that the DQPM results are in agreement with the lattice estimations and close to the range provided by JETSCAPE. For a reasonably good description of the QGP it is important to have 
$\eta/s  > (\eta/s)_{\text{KSS}} = 1/(4\pi)$. 

Comparing results for different setups we see that estimates from setup I are in agreement with the standard DQPM result and with the lattice results, while in the case of setups II and III the values obtained are higher. This can be explained by the smaller masses and widths, resulting from unrealistically small coupling.
The setup I, as well as the original DQPM, is in good agreement with the lattice data for gluodynamics, however, the phenomenologically constrained results from state-of-the-art Bayesian statistical analyses for dynamical, i.e. out of equilibrium, full QCD medium indicate smaller values of $\eta/s$ \cite{Bernhard:2016tnd, Bernhard:2019bmu, Karmakar:2023bxq}. In this respect, we expect that an improved description of the QGP matter should correspond to $\eta/s > (\eta/s)_{\text{KSS}} = 1/(4\pi)$  and within the range of Bayesian estimates. 

Fig. \ref{fig:sigmas-loss-0} presents the scaled electric conductivity, $\sigma_Q/T$, for all flavors, versus the scaled temperature, $T/T_c$. The DQPMnn results derived from Eq. (\ref{eq: sigma_Kubo}) are characterized by red solid lines (representing calculations using $\mathcal{L}_0$ in the setup I), blue solid lines (setup II), and green dashed lines (setup III).
Lattice QCD data points are represented using various symbols: for $N_f=2$, black squares denote the results from Ref. \cite{Brandt:2015aqk}, while for $N_f=2+1$, brown circles represent estimates from Ref. \cite{Aarts:2014nba}, and blue triangles and stars from Ref. \cite{Astrakhantsev:2019zkr}.

The bottom panel of Fig. \ref{fig:sigmas-loss-0} focuses specifically on the scaled electric conductivity of strange quarks, maintaining a similar structure and representation as the upper plot. One can see the same tendency in the case of different setups, i.e. setups II and III predict a larger value of the conductivities due to the smaller effective coupling. 

The consistent overestimation of $\sigma_Q(\text{strange})/T$ across all setups compellingly suggests the need for revising the masses and widths of strange quarks to achieve closer alignment with the lQCD data. This observation is not merely an anomaly but a clear indication that our current understanding and modeling of the strange quark behavior requires refinement.

Furthermore, the simultaneous pursuit of a more accurate agreement with the lQCD data for both $\sigma_Q(\text{full})/T$ and $\sigma_Q(\text{strange})/T$ demands a thoughtful reevaluation and potential modification of the Ansatz employed in the DQPM.
 In the forthcoming two subsections we will explore possible generalizations of the model, which can shed light on which direction microscopic quantities can be adjusted to improve $\chi_2^S$ and $\sigma_Q(\text{strange})/T$.
 
In summary, we have found that setups II and III, which do not fit the lQCD entropy density, provide a better description of $\chi^B_2(T), \chi^S_2(T)$. However, because of the DQPM Ansatz, these two setups yield an unrealistically small value compared to the estimates from lQCD and the analytical 2/1 loop running constant \cite{Caswell:1974gg, Peshier:2006ah} for $\alpha_s = \underline{g}^2(T)/( 4 \pi)$ at $T > 1.5 T_c$. In particular, one can see that the effective coupling $\underline{g}^2(T)\rightarrow 0$ already at $T \approx 2.5 T_c $. 

In the next section we will relax the form of the Ansatz aiming at a consistent simultaneous description of $s/T^3$, $\chi^B_2$ and $\chi^S_2$ in a quasiparticle model.

%------------------------------------------------------
\section{Modified quasiparticle model DNN $\mathcal{L}_{1}$: extraction of $m_i(T), \gamma_i(T)$, and $g^2(T)$ \label{sec:dnn-gen}}

Now we explore a generalization of the DQPM ansatz considered in the previous section. 
The properties of each quasiparticle are the output of the fully connected NN model depicted in Fig. \ref{fig:schemenn}.
For this purpose we consider 6 quantities as output of the DQPMnn: the coupling constant $\underline{g}^2(T)$, gluon and light quark masses, $\underline{M}_g$,$\underline{M}_l$, and three independent quasiparticle widths $\underline{\gamma}_g$, $\underline{\gamma}_l$, $\underline{\gamma}_s$. The mass of the strange quark is still assumed to be $\underline{M}_s=\underline{M}_l+0.03\text{GeV}$. 

In order to make the extraction of the effective coupling $\underline{g}(T)$ possible, we follow the idea proposed in  \cite{Kaczmarek:2005ui} of absorbing the nonperturbative corrections to the masses out of the effective coupling constant.
In the case of gluodynamics, the screening masses were generalized taking into account non-perturbative corrections in the partonic phase in the following way:
\begin{equation}
    \dfrac{m_D(T)}{T}= A(T)(1+ N_f/6)^{1/2}g(T),
\end{equation}
where $A(T)$ describes non-perturbative corrections. As in Ref. \cite{Kaczmarek:2005ui}, $A(T)$ turns out to be larger than the perturbative limit even at high T.
Based on these observations we modify the effective masses using the following ansatz:
\begin{equation}\label{eq: AG hypothesis}
    \dfrac{\underline{m}_{q/g}(T)}{T}= \underline{A}_{i}(T)\underline{g}(T),
\end{equation}
where $\underline{A}_i$ and $\underline{g}$ are computed from the Neural Network. In the DQPM $A_g = 3/4$, $A_q=1/3$. 

It is important to note that, using the above parametrization, $\underline{g}(T)$ remains relatively similar to the original effective coupling $g^{DQPM}(T)$ whereas the masses can differ. This allows to obtain a better description of thermodynamical quantities while preserving transport coefficients and cross-sections from dramatic changes. The widths $\underline{\gamma}_i(T)$ are left completely free, and no parametrization in terms of $\underline{g}(T)$ is assumed. 
This differentiation in non-perturbative corrections within the masses emphasizes that gluons and quarks may exhibit distinct temperature dependence in their non-perturbative adjustments. Additionally, this distinction leaves more freedom to the DQPMnn to learn and adapt the T-dependence of masses and widths, thus achieving a better fit for the susceptibilities $\chi_2^{B,S}$.

In this case study we are relaxing the DQPM assumptions in order to achieve a better agreement with thermodynamic functions. However, to meet physical requirements such as the asymptotic HTL scaling and the hierarchy between masses and widths, we need to modify the loss function by adding a regularization term.

The loss function used in this section reads:
 \begin{align}
 \mathcal{L}_1= \mathcal{L}_0 + \beta_{reg} \mathcal{L}_{DQPM},
 \label{eq:loss ag}
\end{align}
where the regularization term is:
\begin{align}
     \mathcal{L}_{DQPM} = \sum_{i=g,l,s}\left[\underline{\gamma}_i(T)-\gamma_i^{DQPM}(T)\right]^2  + \nonumber \\
     \left[\underline{A}_g(T) - 3/4 \right]^2 +  \left[\underline{A}_q(T)- 1/3 \right]^2.
\end{align}
The inclusion of $\mathcal{L}_{DQPM}$ acts as a constraint on the neural network, ensuring that the produced outputs stay relatively close to the expectations derived from the DQPM. In order to regulate the influence of $\mathcal{L}_0$ on the predictions, one has to fix $\beta_{reg}$ taking into account the relative contribution for this term in the loss function. In this section, we present two distinct setups for $\beta_{i}$, as given in Table \ref{table:weights L1}. 
%%%%%%%%%%%%%% 2 setups
\begin{table}[h]
\centering
\begin{tabular}{|c|c|}
 \hline
   setups  &  $\beta_{G}:\beta_{L}:\beta_{S} : \beta_{reg}$ \\
 \hline
   A  &  $100$ : $1$ : $1$  : $10^3$ \\
   \hline

   B  &  $500$ : $1$ : $1$ : $10^4$ \\
   \hline
\end{tabular}
   \caption{Considered setups for the modified loss function $\mathcal{L}_1$}
   \label{table:weights L1}
 \end{table}
%%%%%%%%%%%%%%
\begin{itemize}
\item 
In the setup ``A'' the weights were chosen such that all the contributions to the loss function are of the same order of magnitude. The use of a larger $\beta_G$ compared to $\beta_{L,S}$ is due to the larger errors $\Delta s_{lQCD}/T^3$ associated with the lattice data for $s/T^3$. Choosing a $\beta_G$ of order one would result in the NN regarding the entropy as an irrelevant feature and learning only from the susceptibilities. Furthermore, the regularization loss is not enhanced by dividing by the variance, hence using a smaller $\beta_{reg}$ would effectively remove any sizeable effect from the regularization loss. 

\item
The setup ``B'', on the other hand, is such that the entropy and the DQPM regularization term have a bigger role compared to the susceptibilities, yielding an intermediate result between the setup ``I'' of the previous section and the setup ``A''. 
\end{itemize}
Furthermore, in this section we train the DQPMnn multiple times randomizing the initial weight distribution each time. The results of these variations are plotted as shaded areas. 

%%%%%%%%%%%%%%%%%%%%%%%%%%%%%%%%%% 3 main EoS
\begin{figure}
\centering
\begin{minipage}[h]{0.95\linewidth}
\includegraphics[width=0.98\linewidth]{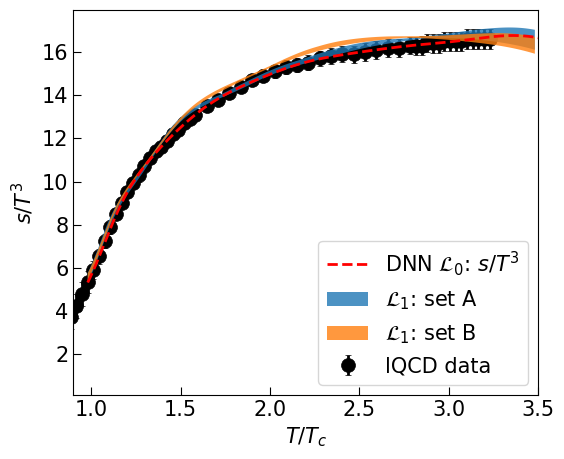}
\end{minipage}
 \begin{minipage}[h]{0.95\linewidth}
\includegraphics[width=0.98\linewidth]{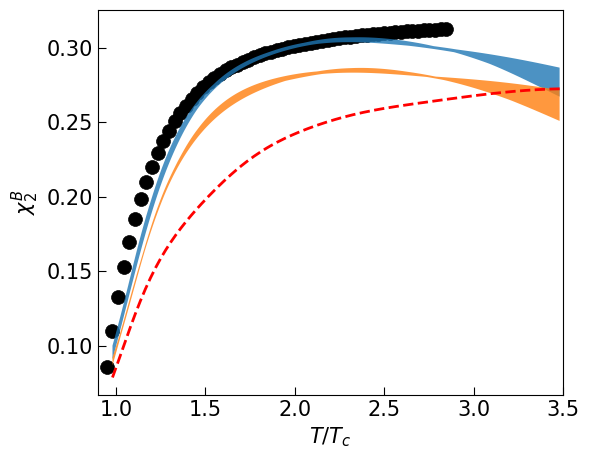} \\
 \end{minipage}
  \begin{minipage}[h]{0.95\linewidth}
\includegraphics[width=0.98\linewidth]{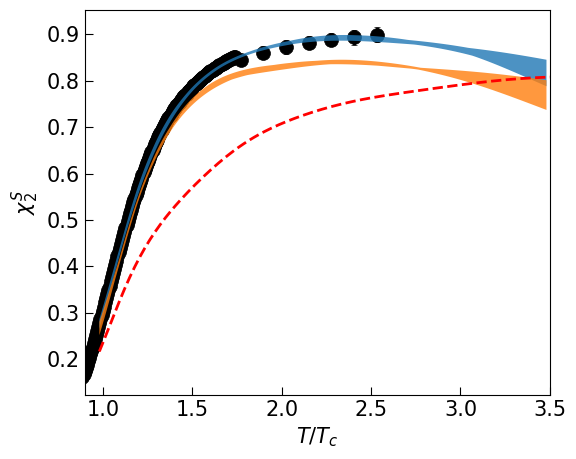} \\ 
 \end{minipage}
\caption{The dimensionless entropy $s/T^3$ (top panel), baryon susceptibility $\chi_2^B$ (middle panel) and strangeness susceptibility  $\chi_2^S$ (bottom panel) as a function of the scaled temperature $T/T_c$. Colored areas correspond to the predictions from the DQPMnn with the modified loss function $\mathcal{L}_{1}$ within the setups A (blue areas), and B (orange areas). The red dashed lines depict predictions generated by the DQPMnn with loss function $\mathcal{L}_{0}:s/T^3$, as discussed in Section \ref{sec:DQPM g2} and trained in setup I. The symbols correspond to the lQCD results from the WB Collaboration \cite{Borsanyi:2011sw, Borsanyi:2013bia, Borsanyi:2022qlh}.}
\label{fig: eos-2sets}
\end{figure}
%%%%%%%%%%%%%%%%%%%%%%%%%%%%%%%%%%
Fig. \ref{fig: eos-2sets} shows the resulting thermodynamic observables for the setups A (blue areas), and B (orange areas) as a function of the scaled temperature $T/T_c$: dimensionless entropy $s/T^3$ (top panel), baryon susceptibility $\chi_2^B$ (middle panel), and strangeness susceptibility  $\chi_2^S$ (bottom panel). The red dashed lines depict predictions generated by the DQPMnn with the loss function $\mathcal{L}_{0} :s/T^3$ (Eq. \eqref{eq:loss-0-dqpm}),  as discussed in Section \ref{sec:DQPM g2} and trained in setup I. The symbols correspond to the lQCD results from the WB Collaboration \cite{Borsanyi:2011sw, Borsanyi:2013bia, Borsanyi:2022qlh}.
The dimensionless entropy $s/T^3$ is described well for both setups, while in the case of setup A, the DQPMnn can describe $\chi_2^B$ and $\chi_2^S$ better than the original DQPM or the DQPMnn $\mathcal{L}_{0}:s/T^3$. This is expected, as the setup B is closer to the standard DQPM due to the larger weight of the entropy and regularization terms $\beta_g$ and $\beta_{reg}$ in the loss function. In both setups, due to the regularization term in the loss function, $\chi_2^B$ and $\chi_2^S$ start to approach DQPM asymptotics for $T>2.5 - 3T_c$, as the absence of lattice data leads the DQPMnn to learn only from the regularization loss term. In general, one can introduce other asymptotic terms for high $T$, such as the Stefan-Boltzmann limit. The main objective here is to obtain physically motivated values of microscopic observables - non-vanishing effective masses and widths, therefore we choose the DQPM form. 

%%%%%%%%%%%%%%%%%%%%%%%%%%%%%%% g2
\begin{figure}[h!]
\centering
\includegraphics[width=0.45\textwidth]{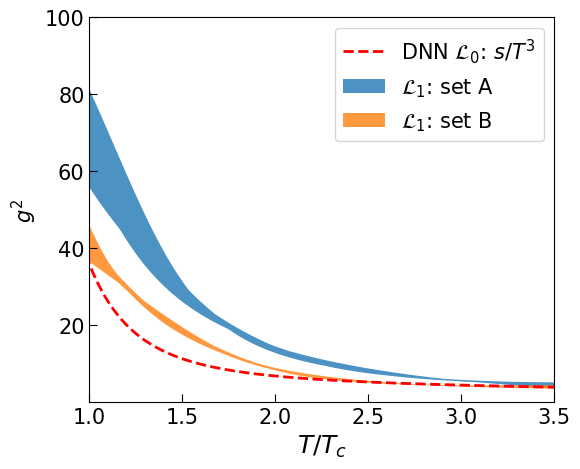}
\caption{The effective coupling $g^2$ as a function of the scaled temperature $T/T_c$ constrained from the DQPMnn with the modified loss function $\mathcal{L}_{1}$ (colored areas) in the setups A-B in comparison to the DQPMnn $\mathcal{L}_{0}:s/T^3$, as discussed in Section \ref{sec:DQPM g2} and trained in the setup I (red dashed line). The color coding for different setups is the same as in Fig. \ref{fig: eos-2sets}.}
\label{fig:ag-g2-2set}
\end{figure}
%%%%%%%%%%%%%%%%%%%%%%%%%%%%%%%%%%

%%%%%%%%%%%%%%% ag, masses, widths
\begin{figure}[h!]
\centering
\includegraphics[width=0.45\textwidth]{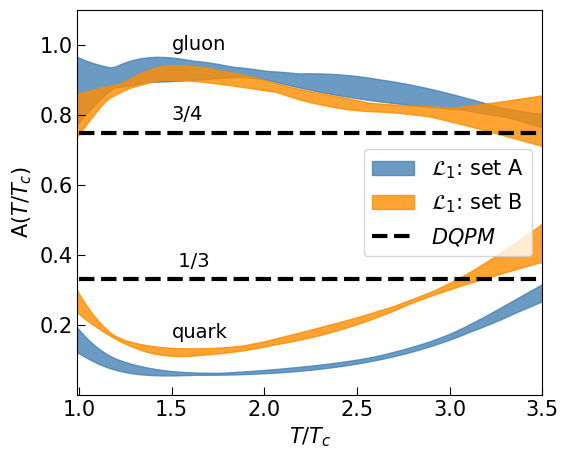}
\caption{The dimensionless quantities $A_q$ and $A_g$ as a function of the scaled temperature $T/T_c$ constrained from DQPMnn with the modified loss function $\mathcal{L}_{1}$ (colored areas) in the setups A-B in comparison to the DQPM values (black dashed lines). The color coding for different setups is the same as in Fig. \ref{fig: eos-2sets}.}
\label{fig:ag-A-2set}
\end{figure}
%%%%%%%%%%%%%%%%%%%%%%%%%%%%%%%%%%
\begin{figure}[!h]
\centering
\begin{minipage}[h]{0.9\linewidth}
\includegraphics[width=0.98\linewidth]{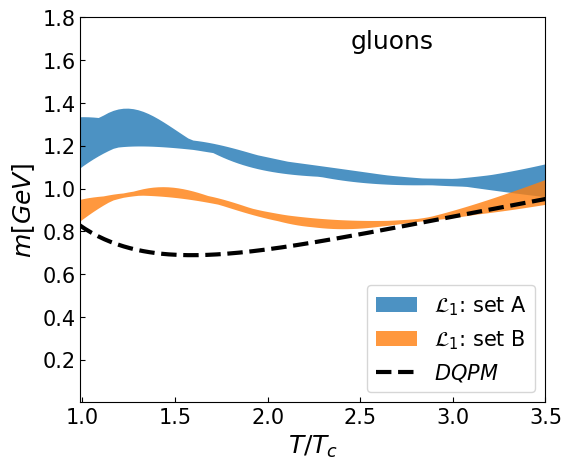}% \\ b) 
\end{minipage}
 \begin{minipage}[h]{0.9\linewidth}
\includegraphics[width=0.98\linewidth]{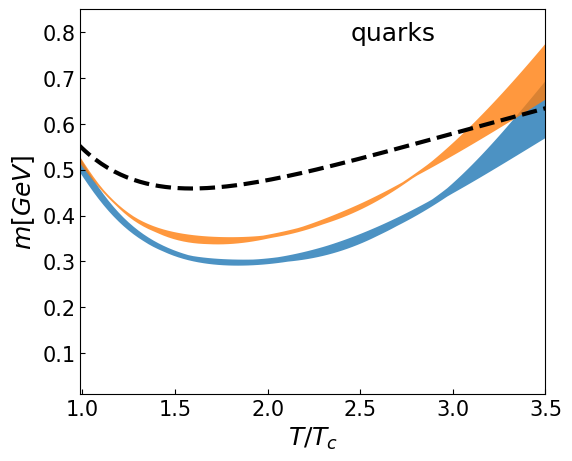}% \\ b) 
 \end{minipage}
 \caption{Masses [GeV] as a function of the scaled temperature $T/T_c$ for gluons (top panel) and light quarks (bottom panel) from the DQPMnn with the modified loss function $\mathcal{L}_{1}$ (colored areas) in the setups A-B in comparison to the DQPM values (black dashed lines). The color coding for different setups is the same as in Fig. \ref{fig: eos-2sets}.}
  \label{fig:vertical-ag-mass-2sets}
    \end{figure}

%%%%%%%%%%%%%%%%%%%%%%%%%%%%%%%%%%
\begin{figure}[!h]
\centering
\begin{minipage}[h]{0.9\linewidth}
\includegraphics[width=0.98\linewidth]{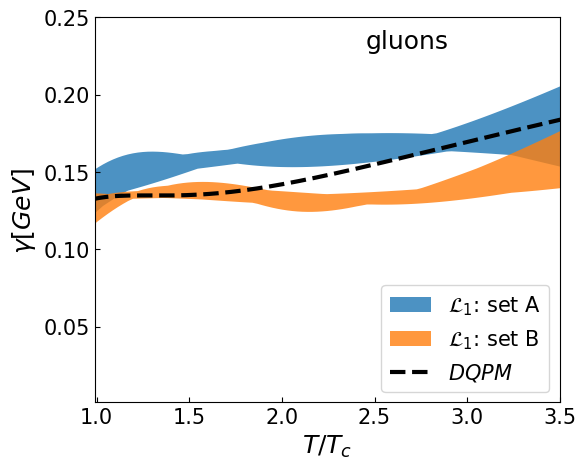}% \\ b) 
\end{minipage}
 \begin{minipage}[h]{0.9\linewidth}
\includegraphics[width=0.98\linewidth]{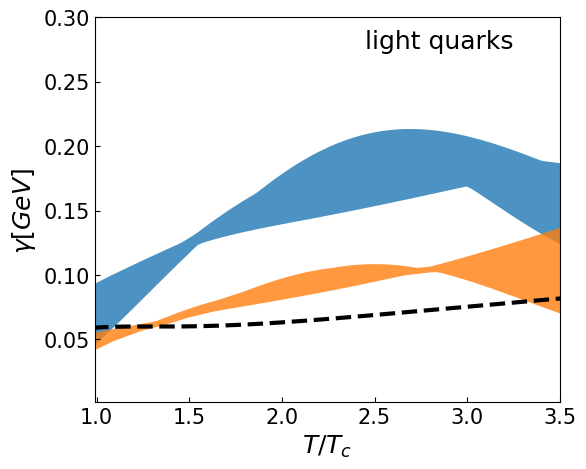}% \\ b) 
 \end{minipage}
  \begin{minipage}[h]{0.9\linewidth}
\includegraphics[width=0.98\linewidth]{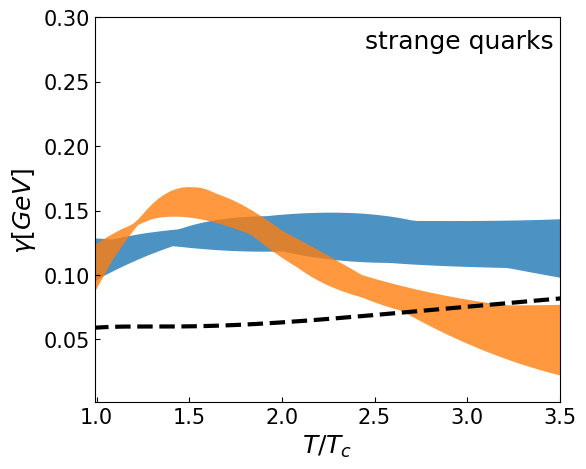}% \\ b) 
 \end{minipage}
 \caption{Widths [GeV] as a function of the scaled temperature $T/T_c$ for gluons (top panel), light quarks (middle panel) and strange quarks (bottom panel) from the DQPMnn with the modified loss function $\mathcal{L}_{1}$ (colored areas) in the setups A-B in comparison to the DQPM values (black dashed lines). The color coding for different setups is the same as in Fig. \ref{fig: eos-2sets}.}

  \label{fig:vertical-ag-wid-2sets}
    \end{figure}
%%%%%%%%%%%%%%%%%%%%%%%%%%%%%%%%%%

%%%%%%%%%%%%%---transport coefficients
%%%%%%%%%%%%%-etas
\begin{figure}[!h]
 \centering
\begin{minipage}[h]{0.985\linewidth}
\includegraphics[width=0.98\linewidth]{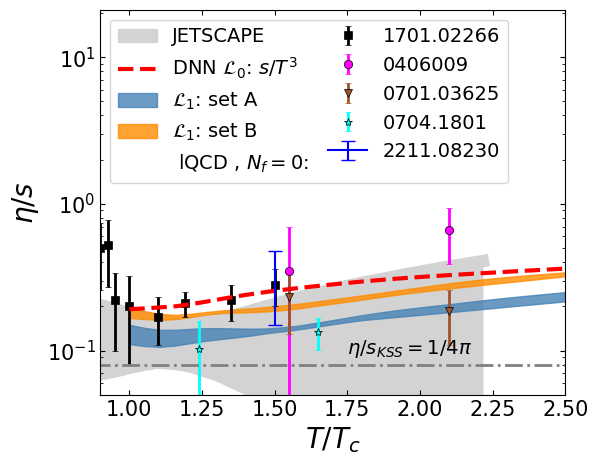}
\end{minipage}
\caption{Specific shear viscosity  $\eta/s$  as a function of the scaled temperature $T/T_c$. Colored areas correspond to estimates from the DQPMnn with the modified loss function $\mathcal{L}_{1}$ within the setups A (blue areas), and B (orange areas) from Eq. (\ref{eta_Kubo}). 
 The red dashed line depicts predictions generated by the DQPMnn with loss function $\mathcal{L}_{0}:s/T^3$, as discussed in Section \ref{sec:DQPM g2} and trained in setup I.
The dashed gray line demonstrates the Kovtun-Son-Starinets bound \cite{Policastro:2001yc,Kovtun:2004de} $(\eta/s)_{\text{KSS}} = 1/4\pi$.  The symbols show lQCD data for pure SU(3) gauge theory taken from Refs. \cite{Astrakhantsev:2017nrs} (black squares), \cite{Altenkort:2022yhb} (blue line),
\cite{Nakamura:2004sy} (green triangles), \cite{Meyer:2007ic} (magenta circles). The grey area corresponds to the latest estimates from a Bayesian analysis by the JETSCAPE Collaboration \cite{JETSCAPE:2020shq}.}
  \label{fig:etas-loss-1}
    \end{figure}

%%%%%%%%%%%%%-sigmas
\begin{figure}[!h]
\centering
\begin{minipage}[h]{0.95\linewidth}
\includegraphics[width=0.98\linewidth]{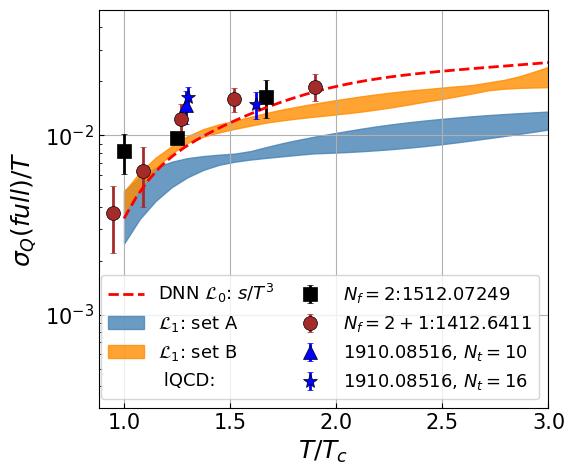}
\end{minipage}
 \begin{minipage}[h]{0.95\linewidth}
\includegraphics[width=0.98\linewidth]{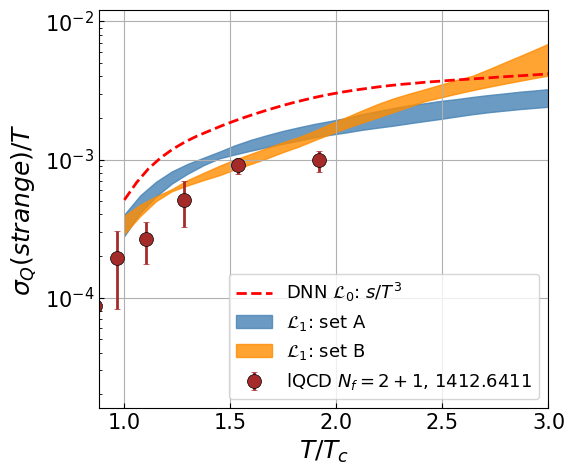} 
 \end{minipage}
\caption{Scaled electric conductivity $\sigma_Q/T$ for all flavors (top panel) and strange quarks (bottom panel) as a function of the scaled temperature $T/T_c$. Colored areas correspond to estimates from the DQPMnn with the modified loss function $\mathcal{L}_{1}$ within the setups A (blue areas), and B (orange areas) from Eq. (\ref{eq: sigma_Kubo}). The symbols correspond to the lQCD results. In the top panel black squares represent $N_f=2$ data from Ref.~\cite{Brandt:2015aqk}, whereas $N_f=2+1$ data are blue triangles and stars from Ref.~\cite{Astrakhantsev:2019zkr}, and brown circles from Ref.~\cite{Aarts:2014nba}. The latter reference is also used for the strangeness contribution to the conductivity in the bottom panel, where lattice data are brown circles.}
  \label{fig:sigmas-loss-1}
    \end{figure}
%%%%%%%%%%%%%%%%%%%%%%%%%%%%%%%%%%
%---------------------------------
\subsection{Results: $g^2, m, \gamma$ and transport coefficients.}
Now we look closely into the resulting microscopic observables and compare them to the original DQPM parametrization. 
Fig. \ref{fig:ag-g2-2set} depicts the effective coupling $g^2$ as a function of the scaled temperature $T/T_c$ constrained in the setups $\mathcal{L}_{1}$: A-B in comparison to the DQPMnn with $\mathcal{L}_{0}:s/T^3$ (setup I, red dashed line). 
In setup A, the coupling exhibits higher values within the $T<3T_c$ range. Unlike setups II and III 
discussed in the previous section, the improved description of $\chi_2^B$ and $\chi_2^S$ in setup A does not lead to unphysically small values of the effective coupling. 
This observation is less surprising considering that, in the setups we are using in this section, macroscopic quantities do not depend solely on $g^2(T)$ but also to the additional $A_i(T)$ parameters, and the widths $\gamma(T)$, which we are leaving completely free. Allowing a general form of the widths,  we can keep the masses of quasi-particles in the physical range of $1- 1.4$ GeV for gluons and $0.3-0.7$ GeV for $T= T_c- 3T_c$.
Examining the dimensionless factors $A_q$ and $A_g$ as functions of the scaled temperature $T/T_c$ (as depicted in Fig. \ref{fig:ag-A-2set})  in setups A-B (colored areas), in comparison to the DQPM values (indicated by the black dashed lines) we can quantify the difference between these scenarios.
Specifically for gluons the observed $A_g$ values are higher than those suggested by the DQPM, in agreement with the results of Ref.~\cite{Kaczmarek:2005ui}. The convergence of these values between the different setups becomes more pronounced for temperatures near $3T_c$, eventually aligning with the DQPM values. This trend is dominantly attributed to the regularization term in the loss function $\mathcal{L}_{1}$. On the contrary, for quarks, the variation among the setups is more pronounced, and the overall values remain consistently lower than those predicted by the DQPM. The smaller value of $A_q$ can be understood from the results about the susceptibilities obtained in the previous section. Indeed, the setups II and III show that, in order to fit $\chi_2^{B,S}$, a smaller coupling is needed. In the previous section, this can also be rephrased by saying that a smaller mass is needed. However, in the present section, the mass is given by the product between $\underline{A}(T)$ and $\underline{g}^2$, therefore a favorable condition to fit the susceptibilities can also be obtained by decreasing the value of $\underline{A}$, as observed in Fig.~\ref{fig:ag-A-2set}.
This underlines a fundamental difference in the behavior of quarks when compared to gluons within the framework provided by the DQPMnn with a relaxed DQPM Ansatz. 

Fig. \ref{fig:vertical-ag-mass-2sets} illustrates masses in GeV as functions of the scaled temperature $T/T_c$ for gluons (top panel) and light quarks (bottom panel).
Here predictions generated by the DQPMnn with $\mathcal{L}_{1}$ are depicted by colored areas within setups A-B contrasted with the DQPM values represented by the black dashed lines. As expected from the results on $g^2$ and $A_{q/g}$ the gluon masses are higher, while quark masses are smaller than the DQPM parametrization. Not imposing the DQPM asymptomatic behavior for $T>3 T_c$ we would expect smaller masses for light quarks and less-pronounced T-dependence for gluons. 

Fig. \ref{fig:vertical-ag-wid-2sets} presents the widths in GeV as a function of the scaled temperature $T/T_c$ across three panels, covering gluons (top panel), light quarks (middle panel), and strange quarks (bottom panel). Similarly to the masses, we show the comparison to the DQPM values represented by the black dashed lines. This comparative analysis reveals the behavior of widths as a function of temperature within these distinct setups depicted by colored areas.
For gluon widths, both setup A and B are close to the DQPM, with setup B showing smaller values. In contrast, for light quarks the overall values are higher than the DQPM, while setup B yields smaller widths. Similarly, for the strange quarks, both setups predict higher values of $\gamma$. Therefore we expect that transport properties will change for setup A and B compared to the DQPM predictions.

Now we look closely into the influence of the change in microscopic quantities on the transport coefficients.
The comparison between DQPMnn results with different loss functions in the case of $\eta/s$ vs $T/T_c$ is depicted in Fig. \ref{fig:etas-loss-1}. The red dashed line shows the predictions generated by the DQPMnn with the loss function $\mathcal{L}_{0}:s/T^3$, as discussed in Section \ref{sec:DQPM g2} and trained in the setup I. 
One can see that for both setups predictions by DQPMnn with the modified description are smaller than in the case of $\mathcal{L}_{0}:s/T^3$. Since the microscopic properties of the setup B are closer to the DQPM parametrization (as have shown above), the $\eta/s$ approaches the value of $\mathcal{L}_{0}:s/T^3$ at $T\approx 2.5 T_c$. The main difference is due to the difference in quark and gluon masses compared to the DQPM values: smaller values of quark masses and increased quark width. In the case of setup A the specific shear viscosity is even smaller due to the larger widths and gluon mass.  

More interesting is to look into the quark sector and compare predictions of DQPMnn with different setups while considering the full and strange electric conductivity. We show the resulting $\sigma_Q/T$, for all flavors (top panel) and strange quarks (bottom panel) as a function of the scaled temperature $T/T_c$ in Fig. \ref{fig:sigmas-loss-1}.
For the total conductivity, we see that predictions from the DQPMnn with $\mathcal{L}_{1}$ in the setup B are larger than setup A, but overall values are smaller than in the case of DQPMnn with $\mathcal{L}_{0}:s/T^3$. In the strange sector, both setups perform better than the standard DQPM model as compared to the lattice data.

%-----------------------------------------------------
\subsection{Strange sector refinements: DNN $\mathcal{L}_{1}$ set A - modifications $\Delta m_{ls}$}

 %%%%%%%%%%%%%%%%%%%%%%%%%%%%%%%%%% 3 main EoS
\begin{figure}
\centering
\begin{minipage}[h]{0.95\linewidth}
\includegraphics[width=0.98\linewidth]{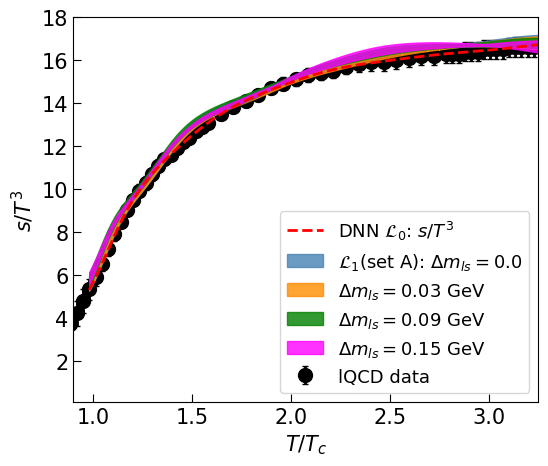}
\end{minipage}
 \begin{minipage}[h]{0.95\linewidth}
\includegraphics[width=0.98\linewidth]{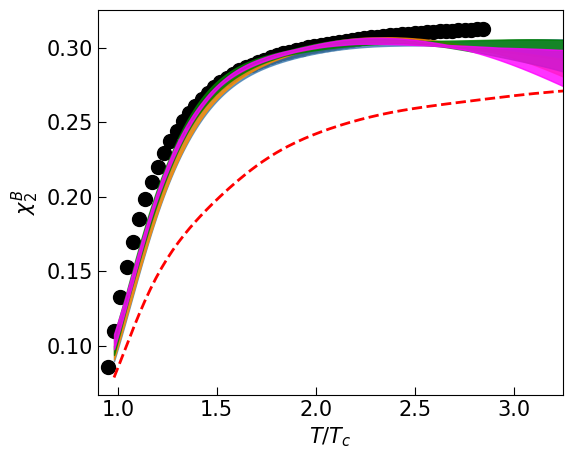} \\
 \end{minipage}
  \begin{minipage}[h]{0.95\linewidth}
\includegraphics[width=0.98\linewidth]{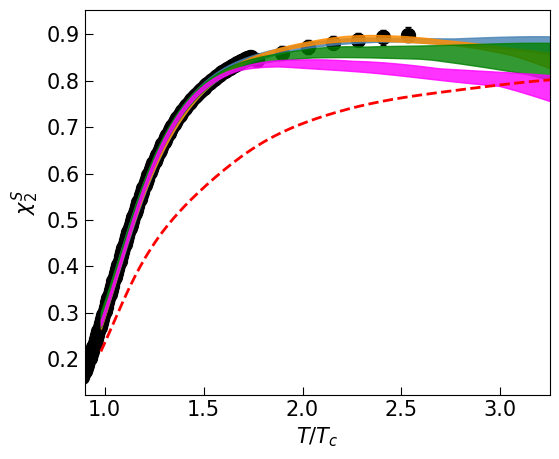} \\ 
 \end{minipage}
\caption{The dimensionless entropy $s/T^3$ (top panel), baryon susceptibility $\chi_2^B$ (middle panel) and strangeness susceptibility  $\chi_2^S$ (bottom panel) as a function of the scaled temperature $T/T_c$. Colored areas correspond to the predictions from the DQPMnn with the modified loss function $\mathcal{L}_{1}$ (set A) with $\Delta m_{ls}$=0.0 GeV (blue areas), $\Delta m_{ls}$=0.03 GeV (orange areas), and $\Delta m_{ls}$=0.09 GeV (green areas), $\Delta m_{ls}$=0.15 GeV (magenta areas). The red dashed lines depict predictions generated by the DQPMnn with loss function $\mathcal{L}_{0}:s/T^3$, as discussed in Section \ref{sec:DQPM g2} and trained in setup I. The symbols correspond to the lQCD results from the WB Collaboration \cite{Borsanyi:2011sw, Borsanyi:2013bia, Borsanyi:2022qlh}.}
\label{fig: eos-all-sets}
\end{figure}

%%%%%%%%%%%%%%% masses, widths
\begin{figure}[!h]
\centering
\begin{minipage}[h]{0.9\linewidth}
\includegraphics[width=0.98\linewidth]{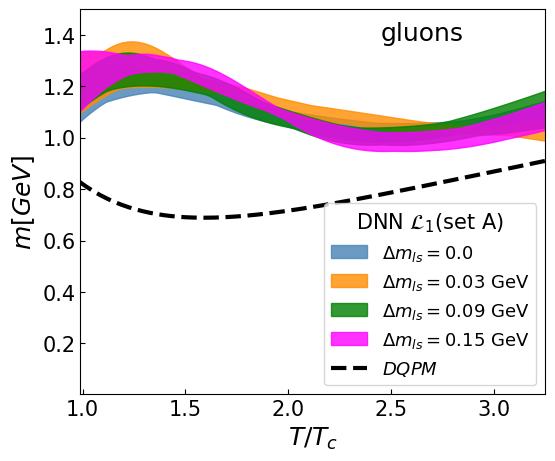}% \\ b) 
\end{minipage}
 \begin{minipage}[h]{0.9\linewidth}
\includegraphics[width=0.98\linewidth]{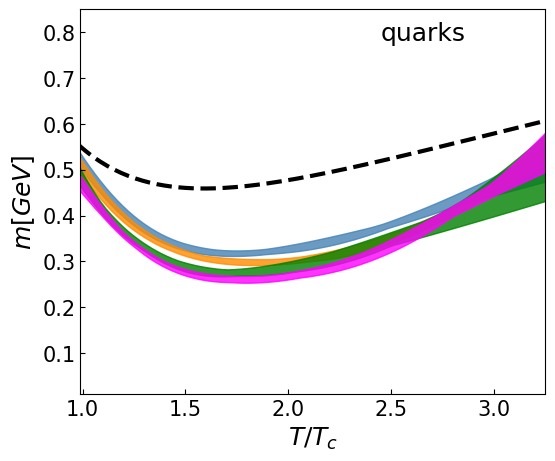}% \\ b) 
 \end{minipage}
 \caption{Masses [GeV] as a function of the scaled temperature $T/T_c$ for gluons (top panel) and light quarks (bottom panel) from the DQPMnn with the modified loss function $\mathcal{L}_{1}$ (set A) with $\Delta m_{ls}$=0.0 GeV (blue areas), $\Delta m_{ls}$=0.03 GeV (orange areas), $\Delta m_{ls}$=0.09 GeV (green areas) and $\Delta m_{ls}$=0.15 GeV (magenta areas) in comparison to the DQPM values (black dashed lines).}

  \label{fig:dms-vertical-ag-mass-2sets}
    \end{figure}

%%%%%%%%%%%%%%%masses
\begin{figure}[!h]
\centering
\begin{minipage}[h]{0.9\linewidth}
\includegraphics[width=0.98\linewidth]{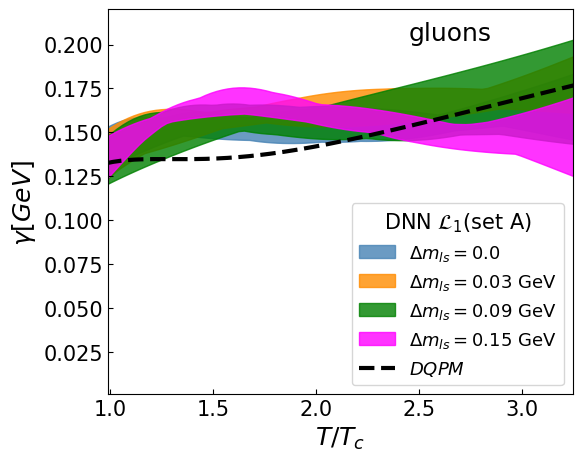}
\end{minipage}
 \begin{minipage}[h]{0.9\linewidth}
\includegraphics[width=0.98\linewidth]{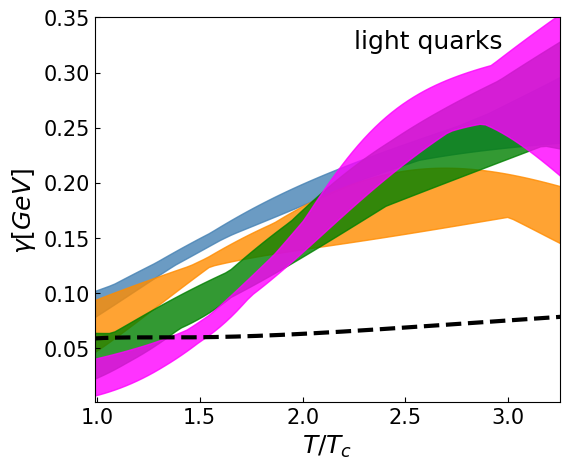}% \\ b) 
 \end{minipage}
  \begin{minipage}[h]{0.9\linewidth}
\includegraphics[width=0.98\linewidth]{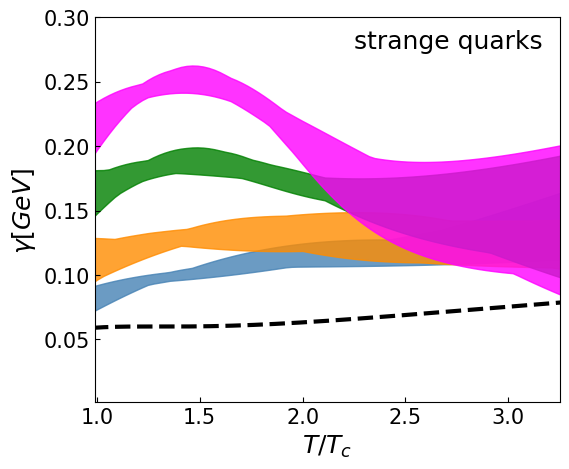} 
 \end{minipage}
 \caption{Widths [GeV] as a function of the scaled temperature $T/T_c$ for gluons (top panel), light quarks (middle panel), and strange quarks (bottom panel) from the DQPMnn with the modified loss function $\mathcal{L}_{1}$ (set A) with $\Delta m_{ls}$=0.0 GeV (blue areas), $\Delta m_{ls}$=0.03 GeV (orange areas), $\Delta m_{ls}$=0.09 GeV (green areas) and $\Delta m_{ls}$=0.15 GeV (magenta areas) in comparison to the DQPM values (black dashed lines).}

  \label{fig:dms-vertical-ag-wid-2sets}
    \end{figure}
%%%%%%%%%%%%%%%

%%%%%%%%%%%%%---transport coefficients
%%%%%%%%%%%%%-etas
\begin{figure}[!h]
 \centering
\begin{minipage}[h]{0.985\linewidth}
\includegraphics[width=0.98\linewidth]{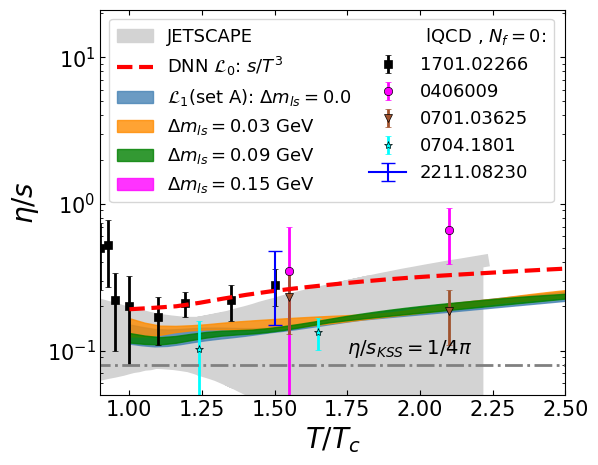}
\end{minipage}
\caption{Specific shear viscosity $\eta/s$  as a function of the scaled temperature $T/T_c$. Colored areas correspond to estimates from the DQPMnn with the modified loss function $\mathcal{L}_{1}$ (set A) with $\Delta m_{ls}$=0.0 GeV (blue areas), $\Delta m_{ls}$=0.03 GeV (orange areas), $\Delta m_{ls}$=0.09 GeV (green areas) and $\Delta m_{ls}$=0.15 GeV (magenta areas) from Eq. (\ref{eta_Kubo}). The red dashed lines depict predictions generated by the DQPMnn with the loss function $\mathcal{L}_{0}:s/T^3$.
The dashed gray line demonstrates the Kovtun-Son-Starinets bound \cite{Policastro:2001yc,Kovtun:2004de} $(\eta/s)_{\text{KSS}} = 1/4\pi$. The symbols show lQCD data for pure SU(3) gauge theory taken from Refs. \cite{Astrakhantsev:2017nrs} (black squares), \cite{Altenkort:2022yhb} (blue line),
\cite{Nakamura:2004sy} (green triangles), \cite{Meyer:2007ic} (magenta circles). The grey area corresponds to the latest estimates from a Bayesian analysis by the JETSCAPE Collaboration \cite{JETSCAPE:2020shq}.}
  \label{fig:dms-etas-loss-1}
 \end{figure}
%%%%%%%%%%%%%-sigmas
\begin{figure}[!h]
\centering
\begin{minipage}[h]{0.95\linewidth}
\includegraphics[width=0.98\linewidth]{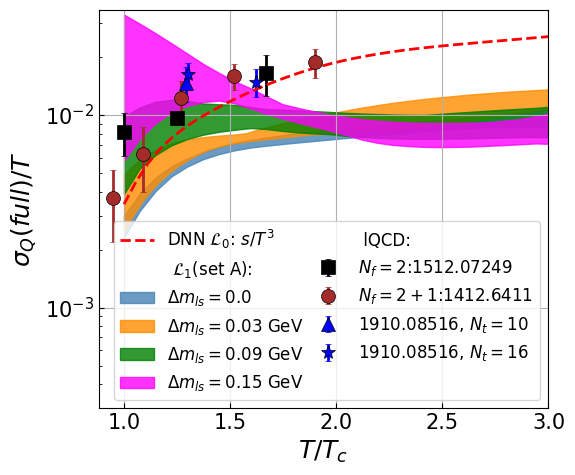}% \\ b) 
\end{minipage}
 \begin{minipage}[h]{0.95\linewidth}
\includegraphics[width=0.98\linewidth]{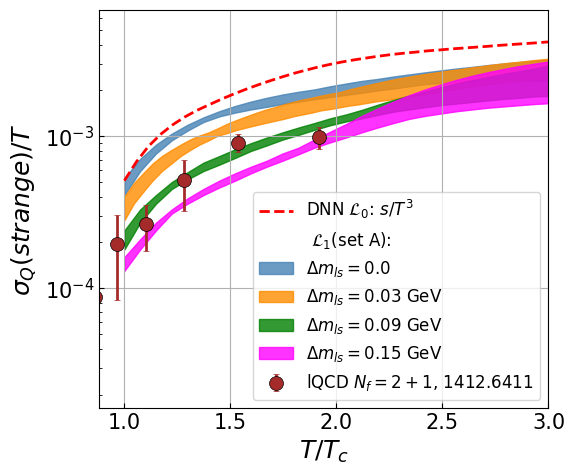}% \\ b) 
 \end{minipage}
\caption{Scaled electric conductivity $\sigma_Q/T$ for all flavors (top panel) and strange quarks (bottom panel) as a function of the scaled temperature $T/T_c$. Colored areas correspond to estimates from the DQPMnn with the modified loss function $\mathcal{L}_{1}$ (set A) with $\Delta m_{ls}$=0.0 GeV (blue areas), $\Delta m_{ls}$=0.03 GeV (orange areas), $\Delta m_{ls}$=0.09 GeV (green areas) and $\Delta m_{ls}$=0.15 GeV (magenta areas) from Eq. (\ref{eq: sigma_Kubo}). The red dashed lines depict predictions generated by the DQPMnn with loss function $\mathcal{L}_{0}:s/T^3$. The symbols correspond to the lQCD results. In the top panel, black squares represent $N_f=2$ data from Ref.~\cite{Brandt:2015aqk}, whereas $N_f=2+1$ data are blue triangles and stars from Ref.~\cite{Astrakhantsev:2019zkr}, and brown circles from Ref.~\cite{Aarts:2014nba}. The latter reference is also used for the strangeness contribution to the conductivity in the bottom panel, where lattice data are brown circles.}
  \label{fig:dms-sigmas-loss-1}
    \end{figure}
%%%%%%%%%%%%%%%%%%%%%%%%%%%%%%%%%%

Here we present results on the influence of $\Delta m_{ls}$ - the difference between the masses of strange and light quarks/antiquarks:
$$\Delta m_{ls} = m_{s(\bar{s})}- m_{l(\bar{l})} .$$  
For this purpose, we train the DQPMnn with the loss function $\mathcal{L}_{1}$ in the setup A for different values of $\Delta m_{ls}$. Now comparing to the standard DQPM Ansatz we have the following modifications for the strange quarks: the value of $\gamma_s$ is not equal to $\gamma_l$, and different values of $\Delta m_{ls}$.
In this section, we present results from four different setups, characterized by $\Delta m_{ls}$ values of $0.0$, $0.03$, $0.09$, and $0.15$ GeV. With these values, we obtain a good fit for the EoS and reasonable values of the transport coefficients. 

If we look into the training observables, $s/T^3$, $\chi_2^{B}$, $\chi_2^{S}$ depicted in Fig. \ref{fig: eos-all-sets} in three panels from top to bottom, respectively, we find that a higher value of  $\Delta m_{ls}$ slightly improves the fit of $\chi_2^B$, however $\chi_2^S$ deviates from the data for $T>2 T_C$. 

Now we look at the microscopic properties predicted by the DQPMnn when we tweak the mass shift $\Delta m_{ls}$.
In Fig.~\ref{fig:dms-vertical-ag-mass-2sets}, we show the masses for the four scenarios, utilizing the $\mathcal{L}_{1}$ loss function (set A) with $\Delta m_{ls}$ set to 0.0 GeV (blue areas), 0.03 GeV (orange areas), 0.09 GeV (green areas) and 0.15 GeV (magenta areas), compared to the DQPM values (depicted by the black dashed line). 
We see that the light-quarks masses are most affected by the mass-shift, becoming smaller as $\Delta m_{ls}$ increases, especially at $T<2T_c$. The gluon mass, on the other hand, stays approximately the same.

Similarly, for the widths shown in Fig. \ref{fig:dms-vertical-ag-wid-2sets} we see that the main deviations between the different scenarios are seen for the quarks. 
The plot displays widths [GeV] as functions of the scaled temperature $T/T_c$ for gluons (top panel), light quarks (middle panel), and strange quarks (bottom panel), derived from the DQPMnn.
We can see that, as $\Delta m_{ls}$ increases, the light-quarks width becomes smaller, whereas the opposite trend is found in the strange-quark width. For $\Delta m_{ls}\geq0.09$ GeV the widths of the strange quarks become larger than the light quarks, and the difference is more prominent for the larger strange mass quark in the region of $T<2.5T_c$, where lattice data for $\chi_2^S$ are present. 
It is noticeable that the asymptotic behaviour of $\gamma_l$ has also changed, and shows different asymptotics for  $T>2 T_c$ compared to the standard DQPM parametrization and predictions from set A.

Now let's look how the mass shift $\Delta m_{ls}$ changes the transport properties.
We start with the shear viscosity, which is less affected by the strange quark. Fig . \ref{fig:dms-etas-loss-1} depicts the comparison of the specific shear viscosity ($\eta/s$) as a function the scaled temperature ($T/T_c$). Colored areas represent estimates from the DQPMnn with the loss function $\mathcal{L}_{1}$ (set A) for four values of $\Delta m_{ls}$ (0.0 GeV in blue, 0.03 GeV in orange, 0.09 GeV in green and 0.15 GeV in magenta) from  Eq. (\ref{eta_Kubo}).
Only a subtle difference can be seen in the vicinity of $T_c$, where the widths and the masses change more drastically.

In Fig. \ref{fig:dms-sigmas-loss-1} the scaled electric conductivity ($\sigma_Q/T$) is shown for all flavors (top panel) and strange quarks (bottom panel), plotted against the scaled temperature ($T/T_c$) computed from Eq.~\eqref{eq: sigma_Kubo}. 
A more precise description is obtained when using the larger $\Delta m_{ls}$ value, particularly within the temperature range of $T<1.5T_c$, for both full and strange conductivities. Nevertheless, there is a decrease in the value of the full conductivity observed in the $T>2T_c$ region, which can be attributed to the temperature-dependent behavior of $\gamma_l$.

We find that for larger masses of the strange quarks the agreement for the electric conductivity from strange quarks becomes better, however, the full conductivity suffers in that case. 

At this stage, it is unclear which specific value within the range of $\Delta m_{ls}$ = 0.0 - 0.15 GeV would be optimal. It would be more beneficial to consider $\Delta m_{ls} = f(T)$ as an additional output of the DQPMnn, however, in the current framework and without additional input from lattice data or HTL calculations, the results obtained are challenging to interpret physically. If this approach is to be pursued, which would be desirable according to the findings discussed, more data are needed. We expect a $T$ dependent mass shift for the strange quark to be larger than $\Delta m_{ls}=0.03$ GeV (DQPM value) in the region $T < 1.5T_c$ and to be decreasing with T.  

%---------------------------------------------------------------
\section{Conclusions and outlook}\label{sec:conclusions}
In this study, we have addressed the phenomenological problem of extraction of the microscopic properties of the off-shell quasi-particles in the QGP using machine learning techniques based on macroscopic observables measured in lattice QCD.

DNNs are used to learn the relationship between $T$, which serves as an input to the network, and dependent variables, i.e. microscopic observables, that are designed as outputs of the network. In particular, we have addressed a few scenarios for temperature-dependent masses, widths, and effective coupling constants inferred from the DNNs.
 In our framework, we trained our DNNs using three independent
thermodynamic quantities, the entropy density $s/T^3$ and the baryon and strange susceptibilities $\chi_2^{B,S}$ using lQCD data provided by the WB Collaboration. 
The key ingredient is that our simple formulation can provide a playground for fixing the strange quark and light quark sectors and studying the interplay between different sectors by utilizing both thermodynamic and transport properties. In addition, it has the flexibility to preserve a chosen scaling by adding additional terms.

Our results cover the temperature range $T_c<T<3.5 T_c$. To assess the quality of our model description of the QGP we have incorporated the Kubo-Zubarev formalism into our framework to compute transport coefficients (the dimensionless specific shear viscosity $\eta/s$ and the ratio of electric conductivity to the temperature $\sigma_Q/T$ ) and compared them to results from first principle lQCD calculations and phenomenologically constrained Bayesian estimations by the JETSCAPE Collaboration. In particular, for the comparison, we consider various lQCD results for the gluodynamics in the case of $\eta/s$, while in the case of the $\sigma_Q/T$ there are recent lattice results for $(2+1)$-flavor QCD that separates additionally the strange quark contribution. 

\textbf{I.} Firstly, we have considered a DNN with one output microscopic quantity, the effective coupling constant $g^2$, and used the DQPM parametrization for quasiparticle's masses and widths. In this framework, we have tested three different setups, labeled I-III,
and explored how model parameters can affect microscopic quantities and transport coefficients. 

We found that setups II and III, despite not fitting the lQCD entropy density, offer a more accurate description of $\chi_2^B(T)$ and $\chi_2^S(T)$. This is possible by using a smaller coupling constant and consequently quark masses are smaller compared to the ones employed in the DQPM. Furthermore, in these setups, the thermal widths are found to be smaller than in the DQPM, which generally translates into a smaller interaction rate and larger transport coefficients (notice however that the interaction rate was not employed in this work, but instead we use the more general Kubo-Zubarev formalism). These observations show that in order to achieve a good simultaneous description of both $s/T^3$ and $\chi_2^{B,S}$, as well as of the transport coefficients, one should reconsider the parameterization for the DQPM. 

\textbf{II.} Such a generalization has been performed in the second part of this work, where we have considered a DNN with 6 output microscopic quantities: temperature-dependent masses, widths, and effective coupling constants. In this case, a regularization term has been introduced in the loss function in order to preserve a HTL-predicted asymptotic behaviour.

The modifications introduced in this section, which have been explored in two setups A and B, allow to fit simultaneously both the entropy density and the susceptibilities. This less constrained model favors heavier gluons and lighter quarks, which is in agreement with the findings of the first part of this paper. For what concerns the thermal widths, the gluon sector is very similar to the DQPM parametrization, whereas the quark widths are larger than in the DQPM. We also observe that the DNN favors different values for the widths of the strange and the light quarks. 

These modifications decrease the value of the transport coefficients. The specific shear viscosity $\eta/s$ now lies within the JETSCAPE predictions. Concerning the conductivities, the description of the strange quark contribution to the conductivity now is in better agreement with the lattice data, however, the total conductivity is found to be smaller than the lQCD results.

\textbf{III.} We further tested another assumption commonly used in the quasi-particle models, namely that the light and strange quark masses are related by a temperature-independent mass shift 
$\Delta m_{ls}$. 
We have explored the value of $\Delta m_{ls}$  in the range $[0,0.15]$~GeV and repeated the previous study, finding that this mass shift also has an impact on transport coefficients. We demonstrate that the strange quark electric conductivity can be better aligned with the lQCD results for a larger $\Delta m_{ls}$. Our study suggests that to achieve a more realistic quark sector description and a better fit for the scaled conductivities considering a temperature-dependent mass shift $\Delta m_{ls}(T)$ with a higher value ($\Delta m_{ls}$>0.03 GeV) near $T_c$ is essential. However, the lack of physical constraints on its value and on its asymptotic scaling, prevents us from making any physically sensible assessment. {In this sense, first-principle-based studies, such as those employing functional methods in the case of Dyson–Schwinger equations (DSE) \cite{Gunkel:2021oya} or functional renormalisation group (FRG) \cite{Gao:2020qsj}, constraining the strange quark mass would be highly desirable.

Overall, our work identifies machine learning as a promising flexible framework, which can provide hints for the improvements of a model description in the phenomenology of HICs. The parameterization of the quasiparticle properties inferred in this work is useful for transport approaches that incorporate a partonic phase, such as the Parton-Hadron-String Dynamics (PHSD), based on Kadanoff-Baym off-shell dynamics (cf. the reviews \cite{Cassing:2008nn, Linnyk:2015rco, Bleicher:2022kcu}). Furthermore, a similar scheme to the one proposed here can be used to study systems at finite baryon density, with a modest chemical potential. In this context, reliable lattice results are being produced, and the generalization capabilities of neural networks may provide yet another tool to investigate the QGP at large $\mu_B$.

%---------------------------------------------
\begin{acknowledgments}

The authors acknowledge inspiring discussions with L. Wang, J. Aichelin, E. Grossi, M. Ruggieri, O. Kaczmarek, and R. Pisarski. A.P. would also like to thank L. Anderlini and P. Braccia for useful suggestions. O.S. would like to thank V. Dusiak for the valuable comments. Also we thank to W. Cassing for the critical reading of our manuscript.
Furthermore, we acknowledge support by the Deutsche Forschungsgemeinschaft (DFG, German Research Foundation) through the grant CRC-TR 211 ``Strong-interaction matter under extreme conditions'' -- project number 315477589 -- TRR 211
as well as for the European Union's Horizon 2020 research and innovation program under grant agreement STRONG--2020 -- No 824093.
The computational resources have been provided by the Center for Scientific Computing (CSC) of Goethe University.

\end{acknowledgments}

\section*{Appendix A: Surrogate model approach \label{appex:surrogate}}
This appendix is devoted to the description of the surrogate model used in this work. As described in the main text, the training process of the neural networks involves the calculation of the integrals in Eqs. \eqref{I and chi integrals} appearing in Eq.  \eqref{thermodynamic functions DQPM}. 
A direct numerical calculation of these integrals during training would be too computationally intensive and time-consuming, making the overall training process prohibitively lengthy.
To circumvent this challenge, we employ a surrogate model. 

The primary role of the surrogate is to approximate the results of the aforementioned integrals in a computationally efficient way, allowing the neural network training to proceed faster.
In practice, we have computed numerically the functions $I_B, I_F$, and $\chi_2^i$ appearing in Eqs. \eqref{I and chi integrals} for different values of temperature $T$, mass over temperature $m/T$ and width over temperature $\gamma/T$. The range of these parameters has been chosen to include the physical values of $T$, $m/T$, and $\gamma/T$ for all quasiparticles as a subset. Therefore, we have computed the integrals in the range $155\text{ MeV}\leq T\leq 510\text{ MeV}$, $0.2\leq m/T\leq 10$ and $0.02\leq \gamma/T\leq 6.2$. In terms of the critical temperature $T_c=158$ MeV, the temperature range explored corresponds to $T\in [0.98 T_c, 3.2 T_c]$ approximately. For the numerical integration, we have used Mathematica \cite{Mathematica}, and generated tables, which were then employed in the training of the surrogate models. These tables are formatted to represent data for $s/T^3$:
\begin{equation*}
   T[\text{GeV}] \ \ |  \ \ m / T  \ \ |  \ \ \gamma /T  \ \ |  \ \ I_B [\text{GeV}^3]  \ \ |  \ \ I_F[\text{GeV}^3]
\end{equation*}   
 and for $\chi_2$:
 \begin{equation*}
     T[\text{GeV}]  \ \ | \ \ m/T  \ \ | \ \ \gamma/T  \ \ |  \ \ \chi^q_2 .     
 \end{equation*}
We use the tables to train two neural networks that take as input $T$, $m/T$, and $\gamma/T$ and give as output the functions $\tilde{I}^{F,B}$ and $\tilde{\chi}_2$, respectively. These neural networks are represented schematically in Figure \ref{fig:scheme_surr}. The loss function used to train these DNNs is the mean squared error, explicitly:
\begin{align*}
    \mathcal{L}^{surr}_{s} = &\sum_{j=I,F}(\tilde{I}^j(T,m/T,\gamma/T)-I^j(T,m/T,\gamma/T))^2,\\
    \mathcal{L}^{surr}_{\chi_2}=&(\tilde{\chi}_2^q(T,m/T,\gamma/T)-\chi_2^q(T,m/T,\gamma/T))^2.
\end{align*}
The surrogates are trained for $6\times 10^4$ epochs and at the end of the training the loss function is $\mathcal{L}^{surr}\sim 10^{-7}-10^{-8}$ for both training and validation samples in both models. The small value of the loss function testifies to the good accuracy achieved by the surrogates in approximating the numerical values of the integrals in Eqs. \eqref{I and chi integrals}.

The surrogate models effectively perform a non-linear interpolation of the data collected in the training table. We prefer to use this approach with respect to standard interpolation procedures because of its speed, its non-linearity, and because of its ability to generalize outside of the range of the table.

The surrogate model is used in the training process as described in the main text. Given the accuracy of the training of the surrogate model and the fact that the physical values of $m/T$ and $\gamma/T$ are well within the range of values used in the training of the surrogate, we expect that numerical errors and artifacts caused by the use of the surrogate will be much smaller than the errors on the lattice data. Therefore, we conclude that the surrogate model can be applied safely to the case studies. Notice that, despite the ability of the neural network to generalize outside of its range of training, any result out of the training intervals of the surrogate models mentioned above is, in principle, less reliable. 

\bibliographystyle{apsrev4-1}
\bibliography{refs} 

\end{document}